\documentclass[authoryear,12pt]{elsarticle}

\usepackage{amssymb}
\usepackage{amsmath}
\usepackage{graphicx}
\usepackage{subcaption}
\usepackage{color}

\journal{Ecological Modelling}

\begin{document}
\begin{frontmatter}
\title{Facultative predation can alter the ant--aphid population}
 \author[label1]{Atsuki Nakai}
 \ead{nakai.atsuki.mb@gmail.com}
 \author[label2]{Yoko Inui}
 \ead{inui@cc.osaka-kyoiku.ac.jp}
 \author[label1]{Kei Tokita\corref{cor1}}
 \ead{tokita@i.nagoya-u.ac.jp}
 \address[label1]{Department of Complex Systems Science, Graduate School of Informatics, Nagoya University, Nagoya 464-8601, Japan}
 \address[label2]{Department of Arts and Sciences, Osaka Kyoiku University, Kashiwara, Osaka 582-8582, Japan}
 \cortext[cor1]{Corresponding author}

\begin{abstract}
Although ant--aphid interactions are the most typical example of mutualism between insect species, some studies suggest that ant attendance is not always advantageous for the aphids because they may pay a physiological cost. In this study, we propose a new mathematical model of an ant--aphid system considering the costs of ant attendance. It includes both mutualism and predation. In the model, we incorporate not only the trade-off between the intrinsic growth rate of aphids and the honeydew reward for ants, but also the facultative predation of aphids by ants. The analysis and computer simulations of the two-dimensional nonlinear dynamical system with functional response produces fixed points and also novel and complex bifurcations. These results suggest that a higher degree of dependence of the aphids on the ants does not always enhance the abundance of the aphids. In contrast, the model without facultative predation gives a simple prediction, that is, the higher the degree of dependence, the more abundant the aphids are. The present study predicts two overall scenarios for an ant--aphid system with mutualism and facultative predation: (1) aphids with a lower intrinsic growth rate and many attending ants and (2) aphids with a higher intrinsic growth rate and fewer attending ants. This seems to explain why there are two lineages of aphids: one is associated with ants and the other is not.
\end{abstract}

\begin{graphicalabstract}
\\
\\
\\
\\
\includegraphics[scale=0.3]{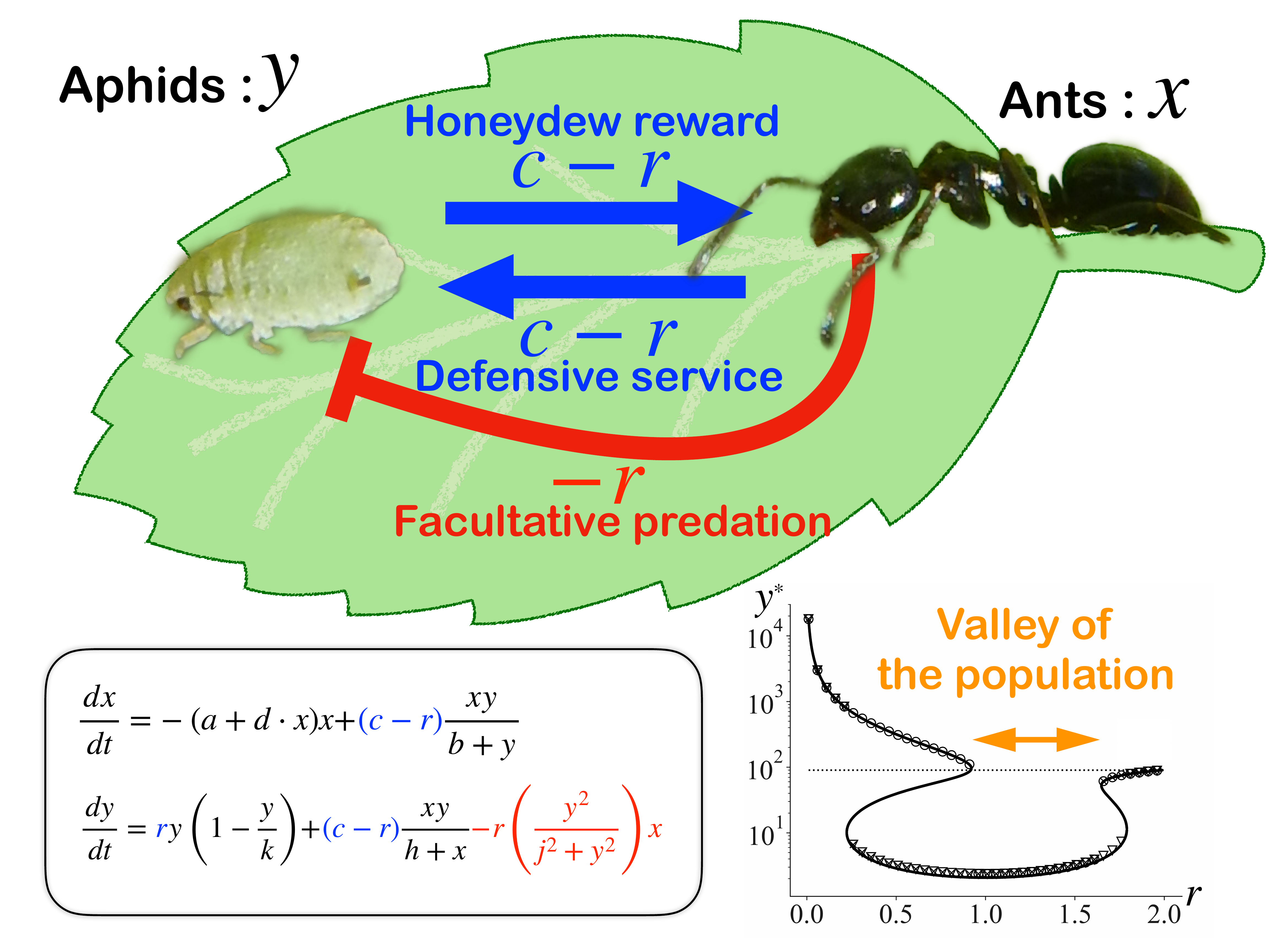}
\end{graphicalabstract}
\begin{highlights}
\item Incorporating the facultative predation of aphids by ants results in a new understanding of mutualism. 
\item A moderate dependence of the aphids on ants increases the aphid extinction rate.
\item Aphids do not require single-minded attendance by ants.
\item The mathematical model predicts that there should be two lineages of aphids: those with and those without ants. 
\item Facultative predation may be an example of a Holling's type III functional response.
\end{highlights}
\end{frontmatter}

\section{Introduction}
The ant--aphid interaction, one of the most typical examples of mutualism, has been actively researched by field ecologists.
Ants harvest the honeydew excreted by aphids and, in turn, protect the aphids from predators. In addition, since excessive honeydew, which is excrement for the aphids, degrades the aphid's habitat, the consumption of honeydew by ants is also beneficial to aphids since it prevents such environmental deterioration  \citep{nixon1951association,nielsen2010ants}.
However, there is a theory that aphids pay a physiological cost in producing the high-quality honeydew needed to attract ants \citep{stadler1998costs,yao2000costs,yao2014costs}.
In addition, it has been reported that attending ants prey on aphids when the aphid density per ant is high \citep{sakata1994ant,sakata1995density}. Moreover, there are aphid species attended by few or no ants \citep{Bristow1991}. 

On the other hand, the history of mathematical models for mutualism is not very long compared to those for predation or competition.  The classical model started with a simple extension that reversed the sign of the species interaction in the Lotka--Volterra competition system \citep{Vandermeer1978}. This model, however, is not realistic because the population can explode depending on the value of a parameter. To prevent such a population explosion, a functional response term was introduced into the model \citep{Wright1989}. This was the first realistic model of mutualism but it focused only on the benefit of mutualism.
In contrast, from the beginning of this century, some studies have considered the cost paid by the mutualist as well as the benefit. Such models are referred to as consumer--resource interaction models and they incorporate the cost into the functional response term \citep{Holland2002,Holland2010}.

In this paper, we propose a new mathematical model for ant--aphid systems. It incorporates the trade-off between the intrinsic growth rate of aphids and the honeydew reward for ants. It is based on the biological insight that aphids allocate some of their available resources to produce high-quality honeydew \citep{yao2000costs}.
In addition, in ant--aphid systems, it is known that ants prey on aphids if the aphid density per ant exceeds a certain value or if the quality of the honeydew reduces \citep{sakata1994ant,sakata1995density}.
The main purpose of the present study is to clarify the significance of such facultative predation, since it has not previously been discussed mathematically.

\begin{figure*}[th]
\centering
\includegraphics[scale=0.35]{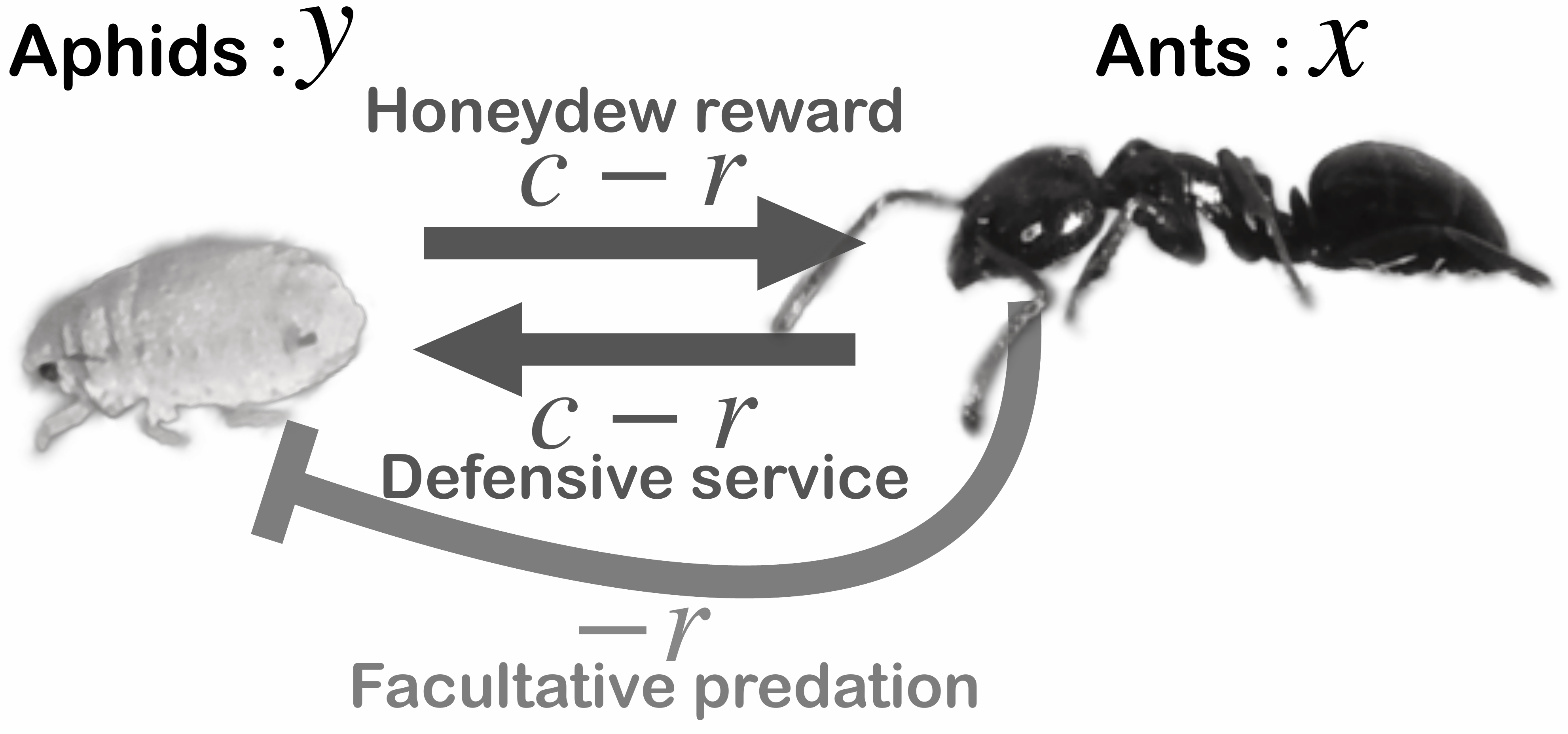}
\caption{Mutualistic relationships between aphids and ants, and the facultative predation of aphids by ants.}
\label{Fig1}
\end{figure*}

\section{Model}
Based on the above discussion, the mathematical model considered in the present study is as follows:
\begin{align}
\frac{dx}{dt}&=-D(x)x+\left \{c'-f(r)\right \}\left.\frac{xy}{b+y}\right.,\label{Eq1}\\
\frac{dy}{dt}&=ry\left(1-\frac{y}{k}\right)+m\left \{c'-f(r)\right \}\frac{xy}{h+x}-H(r)\left.\left(\frac{y^2}{j^2+y^2}\right)x\right.,\label{Eq2}
\end{align}
where $x$ and $y$ are the ant and aphid populations on the host plant, respectively. 
The parameters $c'$, $r$, and $k$ denote, respectively: (1) the total amount of resource consumed by an aphid and used for reproduction, (2) the intrinsic growth rate of aphids including death by predators such as ladybirds, and (3) the carrying capacity for the aphids. 
In general, the resource that aphids allocate to their self-reproduction is represented by a function of $r$, $f(r)$, and the balance $c'-f(r)$, therefore, denotes the amount of resource that aphids allocate to producing honeydew, which is the trade-off between the intrinsic growth rate of aphids and the honeydew reward for ants. 
The second terms of the right-hand sides of  Eqs.~(\ref{Eq1}) and (\ref{Eq2}) represent the mutualistic interaction expressed by the Holling's type II 
functional response, which has been used in models in other studies \citep{Wright1989,Holland2002}.
Mathematically, the nonlinear parameters $b$ and $h$ are the half-saturation populations for aphids and ants, respectively.
In the context of entomology, the parameters $b$ and $h$ can be expressed as
\begin{align}
b &= \frac{1}{e_{yx}t_{h}},\label{Eq3}\\
h &= \frac{1}{e_{xy}t_{a}},\label{Eq4}
\end{align}
where $e_{yx}$, $e_{xy}$, $t_{h}$, and $t_{a}$ denote the rate at which an ant encounters aphids, the rate at which an aphid encounters ants,
the average handling time by ants, and the average time an ant spends attending aphids, respectively.
The parameter $m$ is introduced because the same amount of resources contributes differently to the growth of ants (syntrophy) and aphids (defensive service). 
Note that $x$ is the population of ants on the aphid's host plant and that ants sometimes return to their nest.
In general, such a homing rate (which includes the death rate) of ants is described by the function $D(x)$. 
The facultative predation of aphids by ants, 
the third term on the right-hand side of Eq.~(\ref{Eq2}), is represented by the product of three terms: (1) the predation rate $H(r)$, which is a function of $r$ in general, (2) the Holling's type III functional response $y^2/(j^2+y^2)$, and (3) $x$. 
The nonlinear parameter $j$ is the half-saturation population of aphids. It is similar to $b$, but, in the context of entomology, the accelerating function $y^2/(j^2+y^2)$ is, in general, due to the learning time of ants. We use the type III functional response for predation instead of the type II functional response because previous studies \citep{sakata1994ant,sakata1995density} reported that ants start to prey on aphids when the aphid population exceeds some value. The aphids are a protein source for ant larvae, and the ants even chemically mark aphids for efficient harvesting or predation later.
This type of learned behavior is best modeled by a Holling's type III response function for facultative predation.

We now assume that the above functions have simple forms:
\begin{align}
D(x)&=a+d\cdot x,\label{Eq5}\\
H(r)&=h_0+h_1\cdot r,\label{Eq6}\\
f(r)&=f_0 + f_1\cdot r. \label{Eq7} 
\end{align}
We further assume that $ f_0=0$, $f_1=1$, $h_0=0$, $h_1=1$, and $m=1$ for simplicity and define $c\equiv c'-f_0$. Hence, the nonlinear differential equations for the two species that incorporate mutualism, the trade-off between $r$ and $c-r$, and facultative predation are as follows:
\begin{align}
\frac{dx}{dt}&=-\left(a+d\cdot x\right)x+\left (c-r\right )\frac{xy}{b+y},\label{Eq8}\\
\frac{dy}{dt}&=ry\left(1-\frac{y}{k}\right)+\left(c-r\right)\frac{xy}{h+x}-r\left(\frac{y^2}{j^2+y^2}\right)x,\label{Eq9}
\end{align}
where $a$ and $d$ denote the homing rate of ants to their nest and the self-limitation of ants, respectively.
Fig.~\ref{Fig1} is a conceptual diagram of the relationships between aphids and ants.
We assume the parameters $a$, $b$, $c$, $d$, $h$, $j$, $k$, and $r$ are all positive and that $c>r$, so that the second terms on the right-hand sides of Eqs.~(\ref{Eq8}) and (\ref{Eq9}) are mutualistic. 
Note that this model requires the second-order self-regulation term, $-d \cdot x^2$, to avoid a population explosion of ants.

\section{Results}
 \begin{figure*}[h]
 \begin{minipage}[b]{0.5\linewidth}
  \centering
  \includegraphics[scale=0.34,bb=0 0 461 346]
  {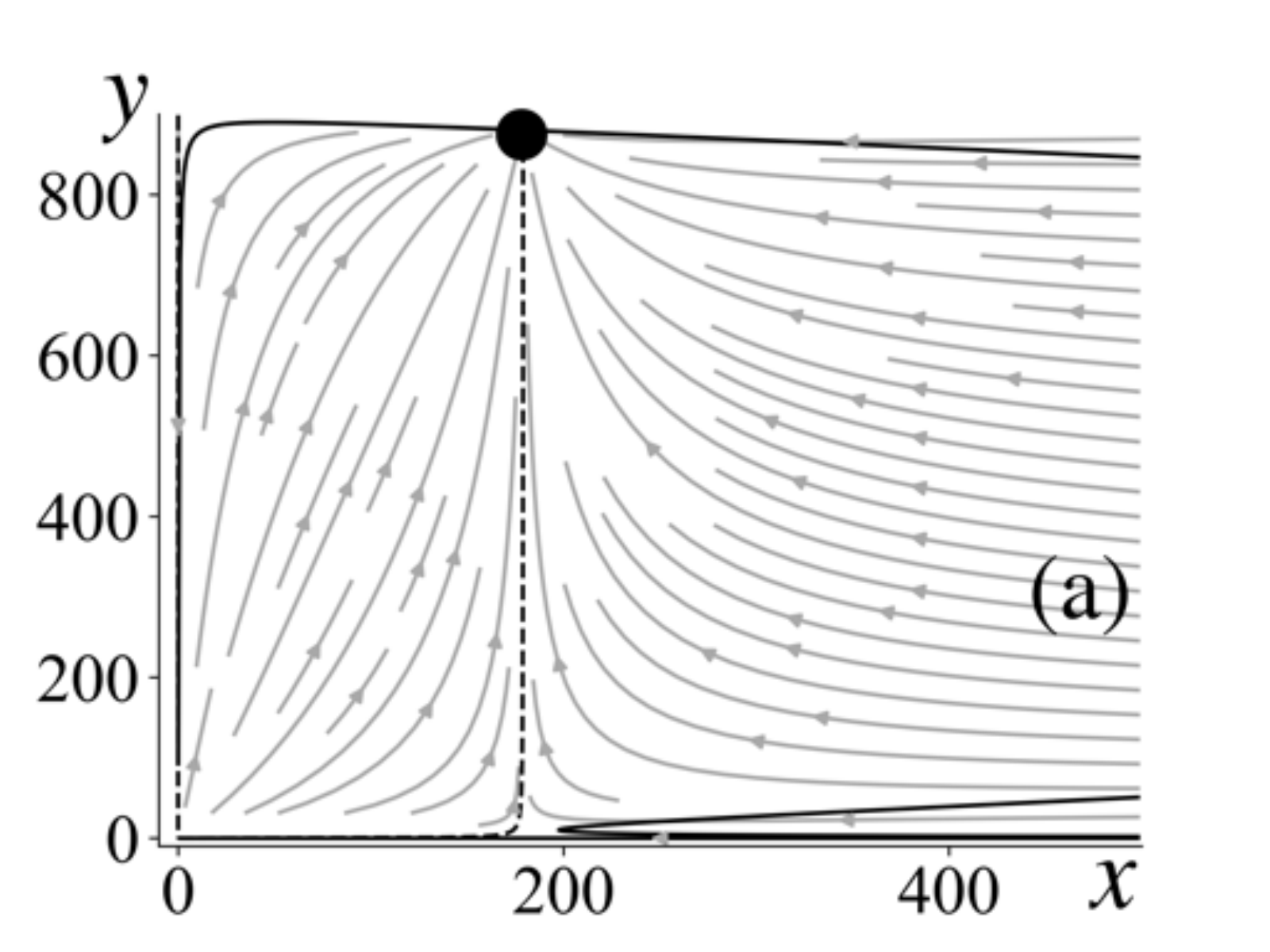}
 \end{minipage}
 \begin{minipage}[b]{0.5\linewidth}
  \centering
  \includegraphics[scale=0.34,bb=0 0 461 346]
  {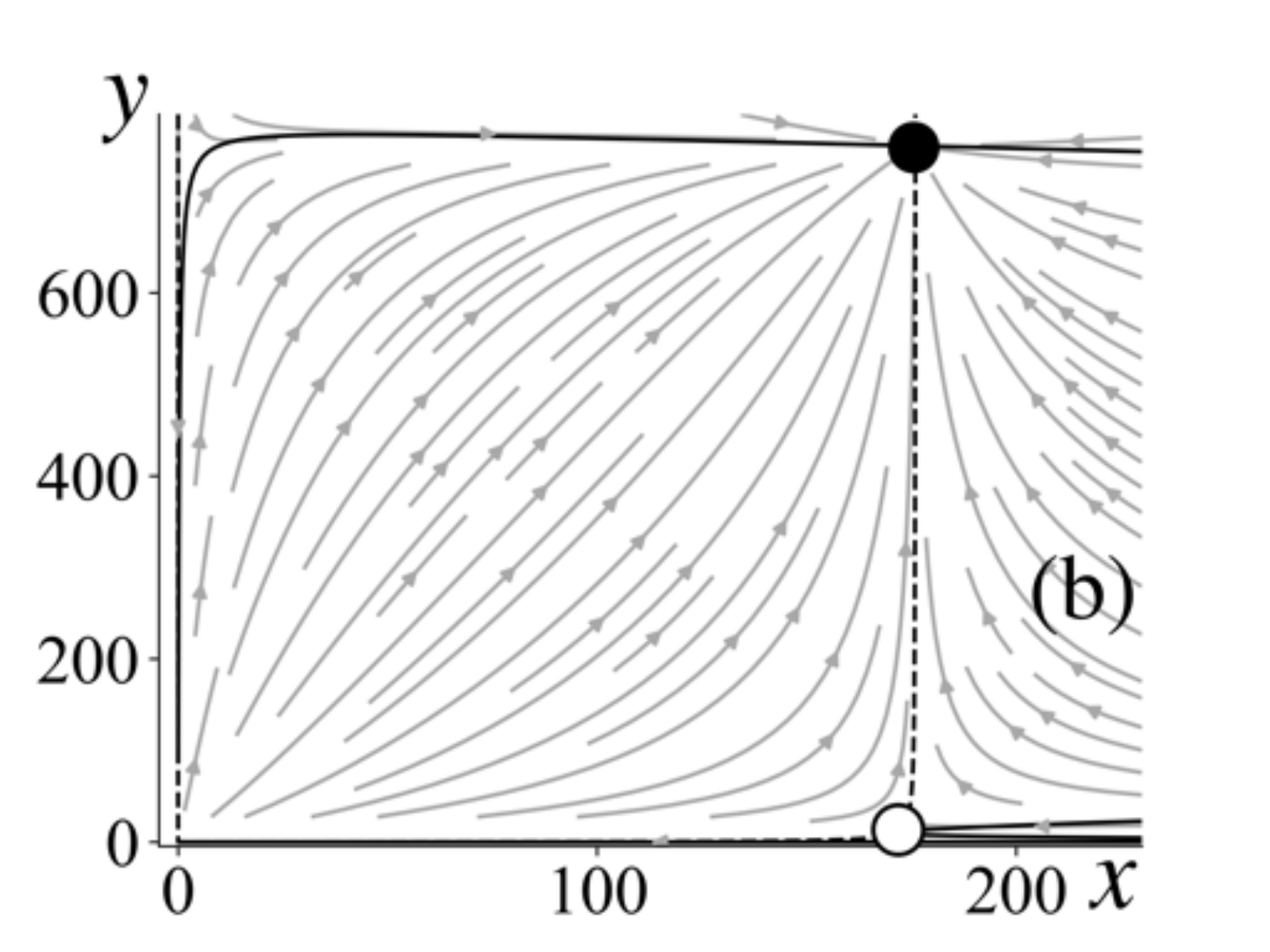}
 \end{minipage}\\
 \begin{minipage}[b]{0.5\linewidth}
  \centering
  \includegraphics[scale=0.34,bb=0 0 461 346]
  {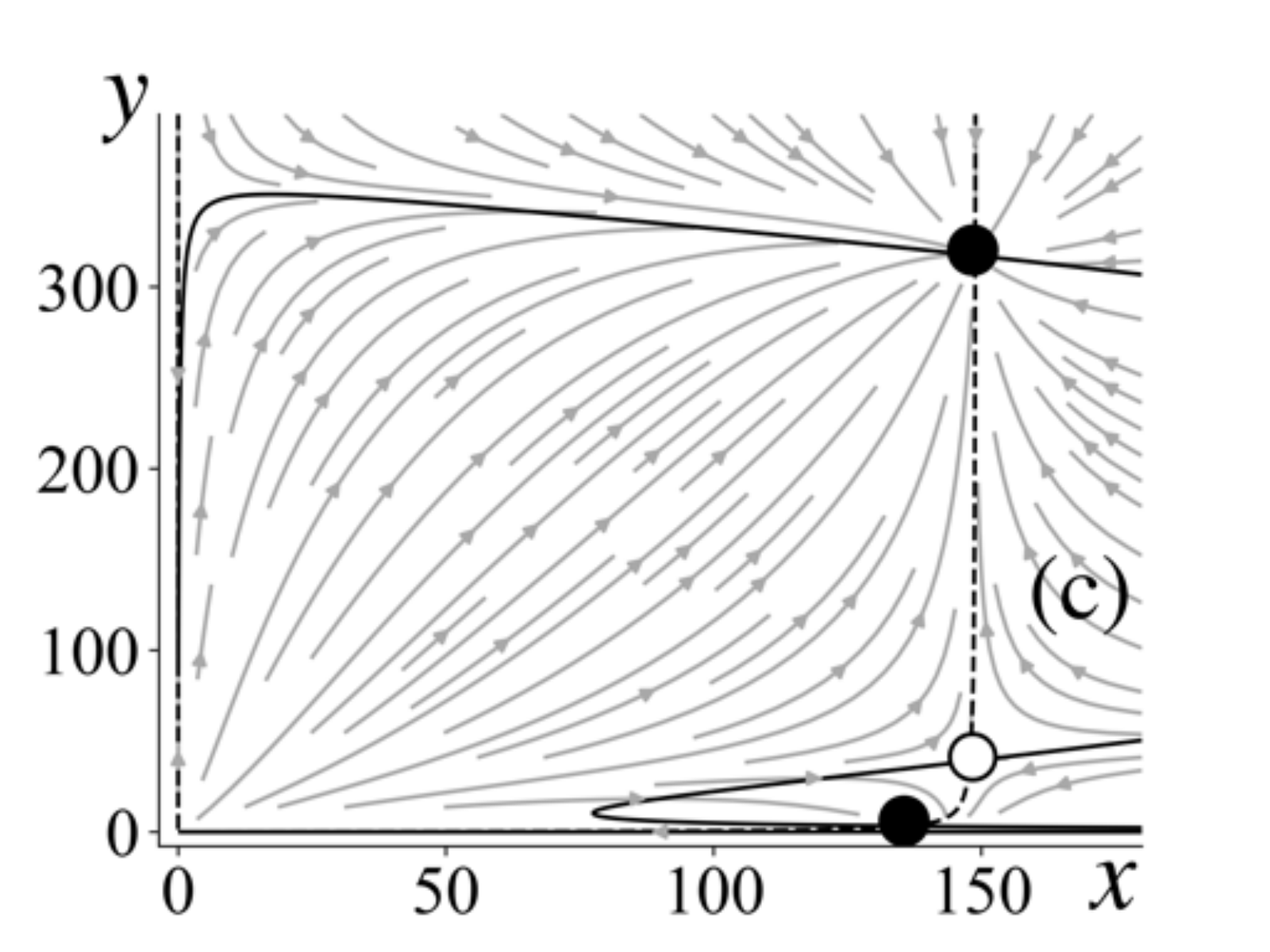}
 \end{minipage}
 \begin{minipage}[b]{0.5\linewidth}
  \centering
  \includegraphics[scale=0.34,bb=0 0 461 346]
  {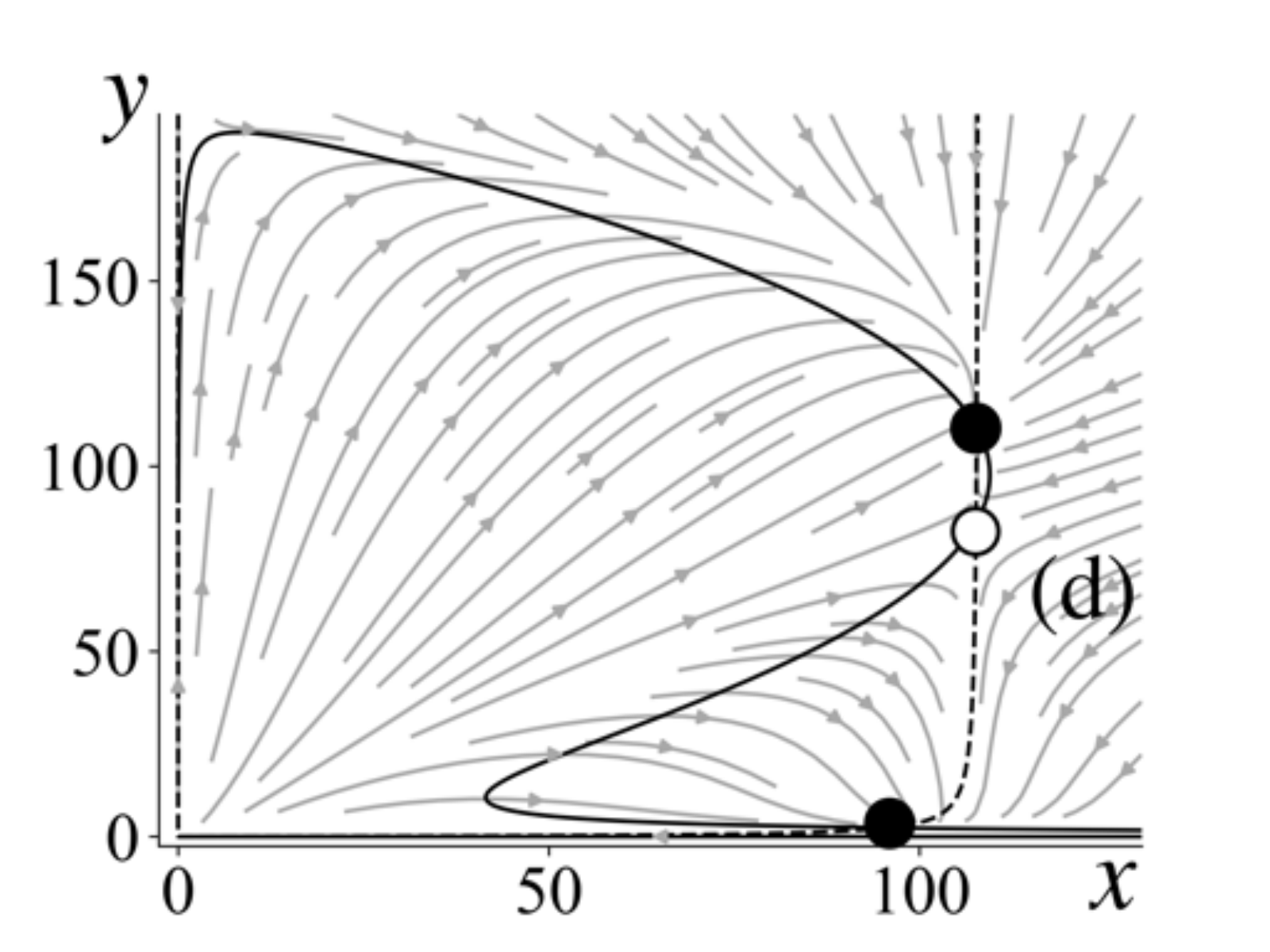}
 \end{minipage}\\
 \begin{minipage}[b]{0.5\linewidth}
  \centering
  \includegraphics[ scale=0.34,bb=0 0 461 346]
  {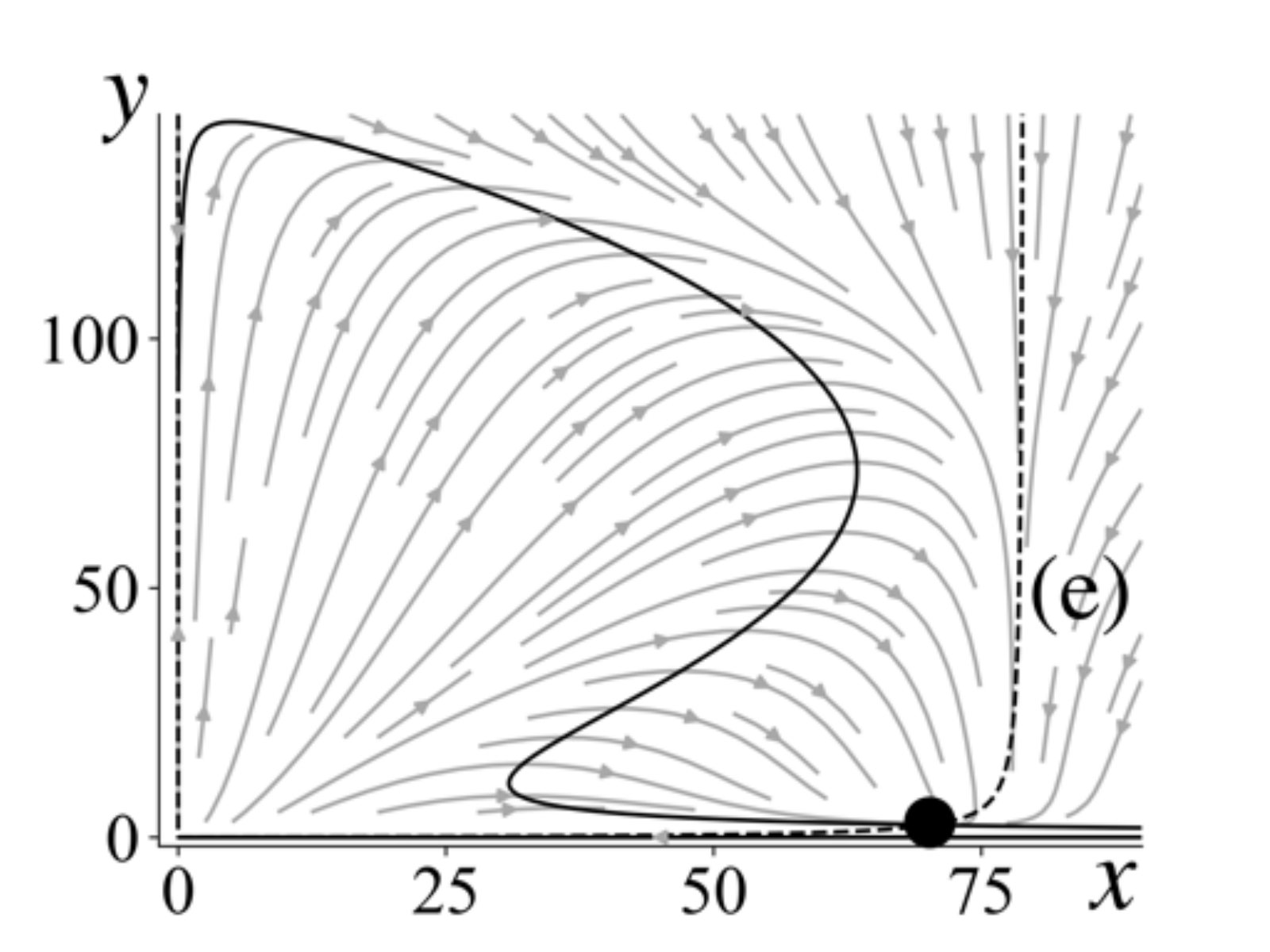}
 \end{minipage}
  \begin{minipage}[b]{0.5\linewidth}
  \centering
  \includegraphics[ scale=0.34,bb=0 0 461 346]
  {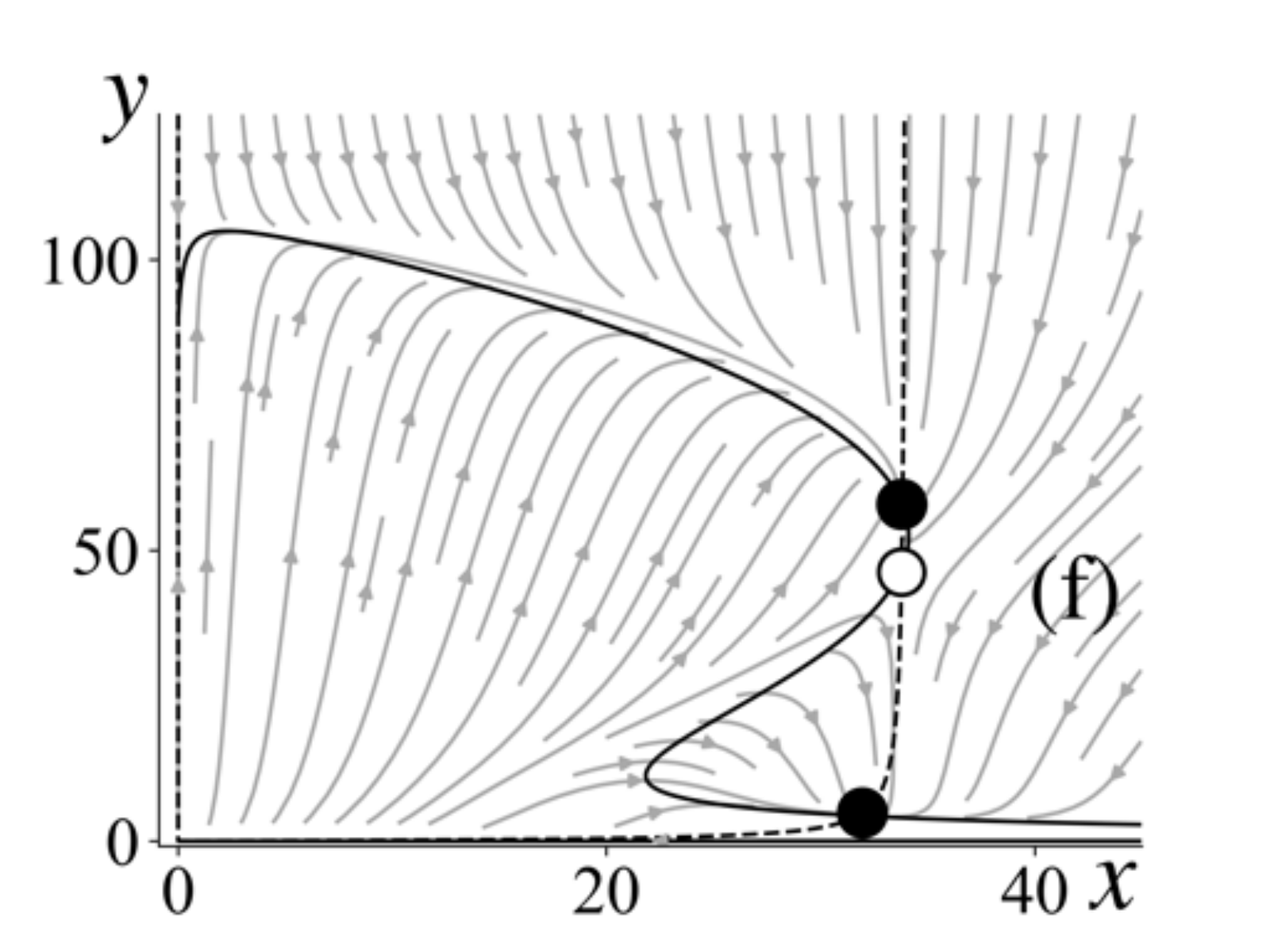}
 \end{minipage}\\
  \begin{minipage}[b]{0.5\linewidth}
  \centering
  \includegraphics[scale=0.34,bb=0 0 461 346]
  {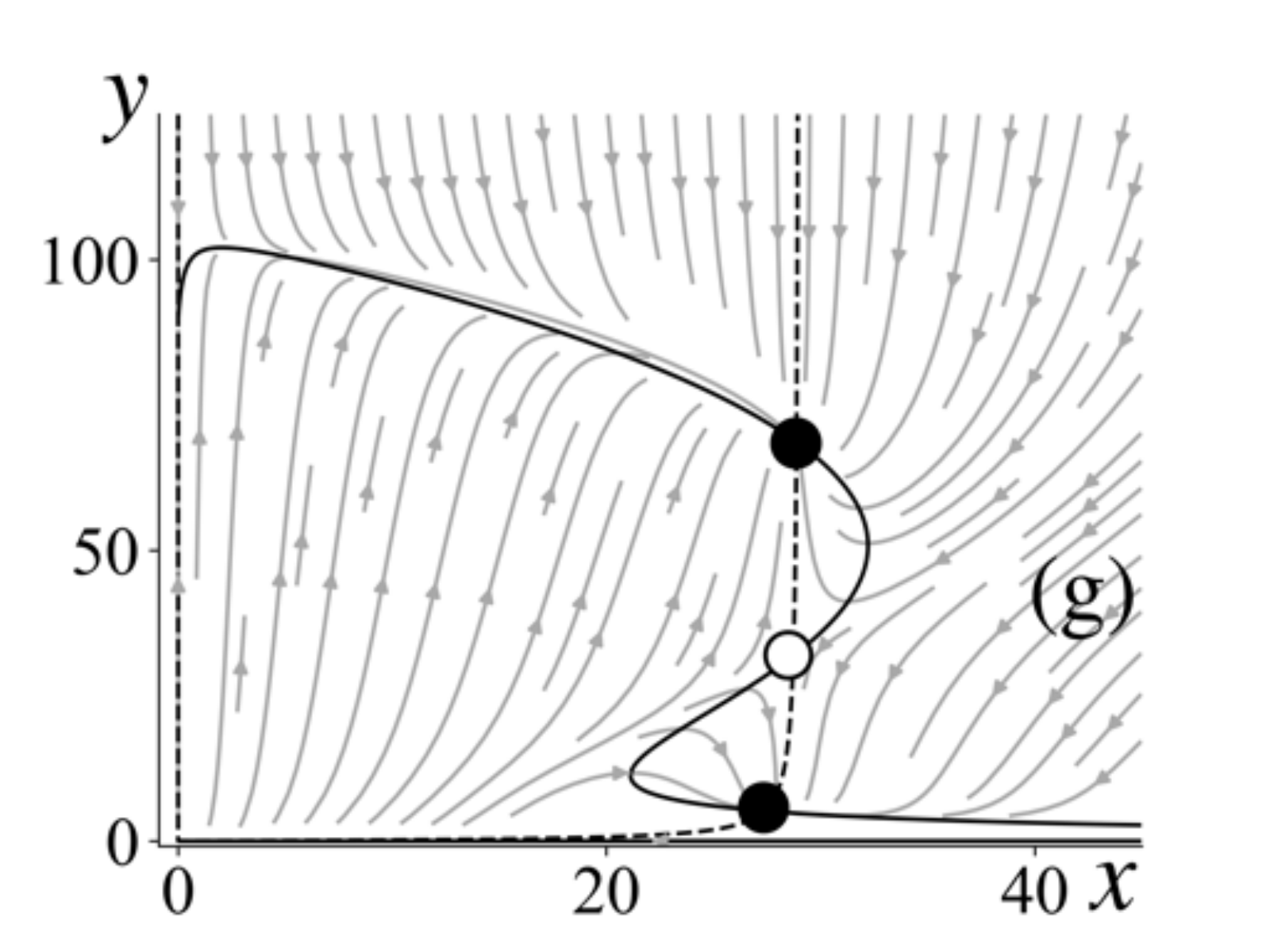}
 \end{minipage}
 \begin{minipage}[b]{0.5\linewidth}
  \centering
  \includegraphics[scale=0.34,bb=0 0 461 346]
  {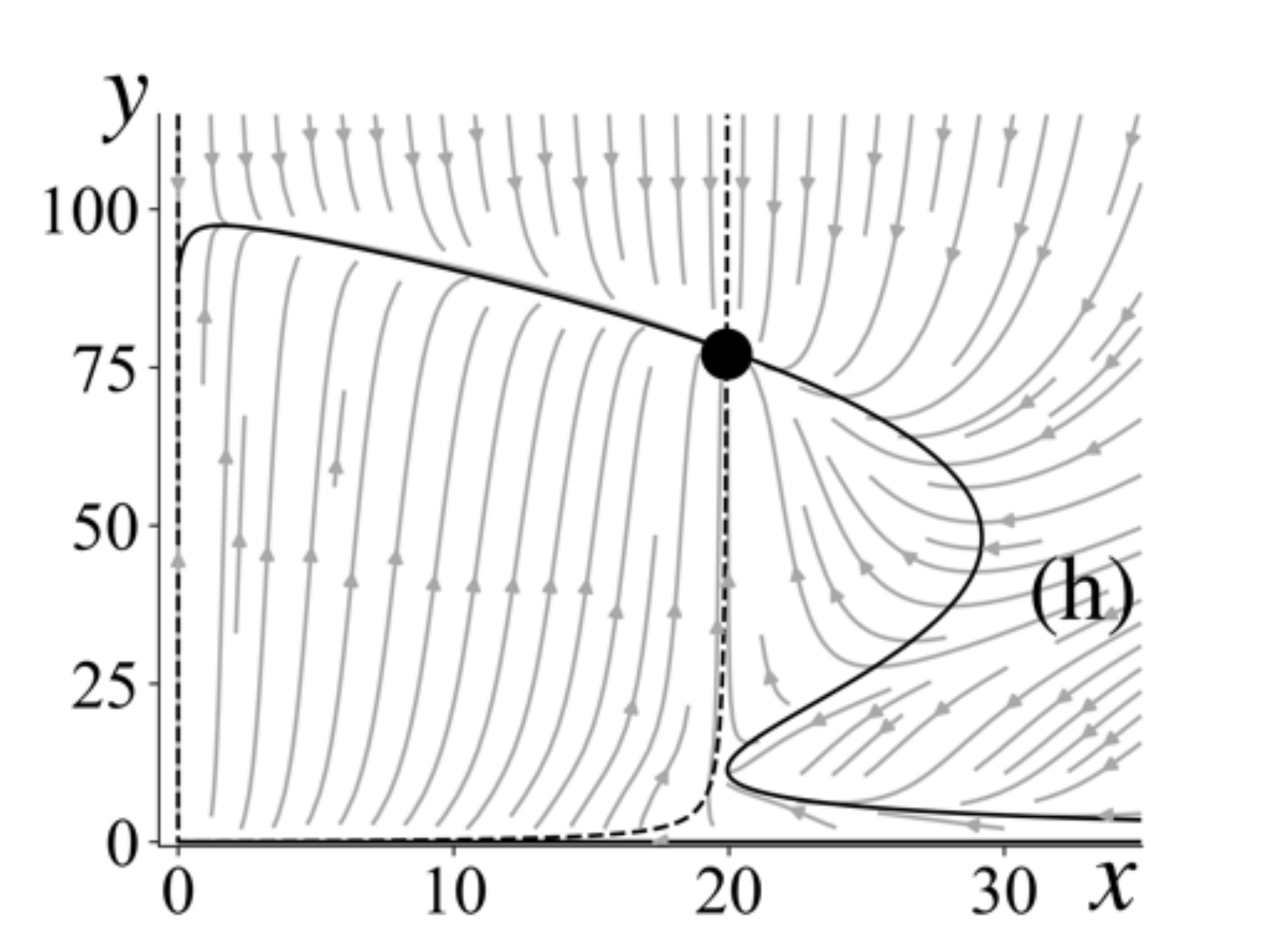}
 \end{minipage}
 \caption{
Flows, with locally stable fixed points (black circles), unstable fixed points (white circles), $x$-nullclines (dashed lines), and $y$-nullclines (solid lines),
for the mutualistic ant--aphid system with facultative predation. The values of the parameters are $a=0.01$, $b=0.3$, $c=2.0$, $d=0.01$, $h=0.3$, $j=10$, and $k=90$, and (a) $r=0.20$, (b) $r=0.23$, (c) $r=0.50$, (d) $r=0.91$, (e) $r=1.20$, (f) $r=1.65$.}

\label{Fig2}
\end{figure*} 

\begin{figure*}[h]
\centering
\includegraphics[scale=0.85,bb=0 0 461 346]{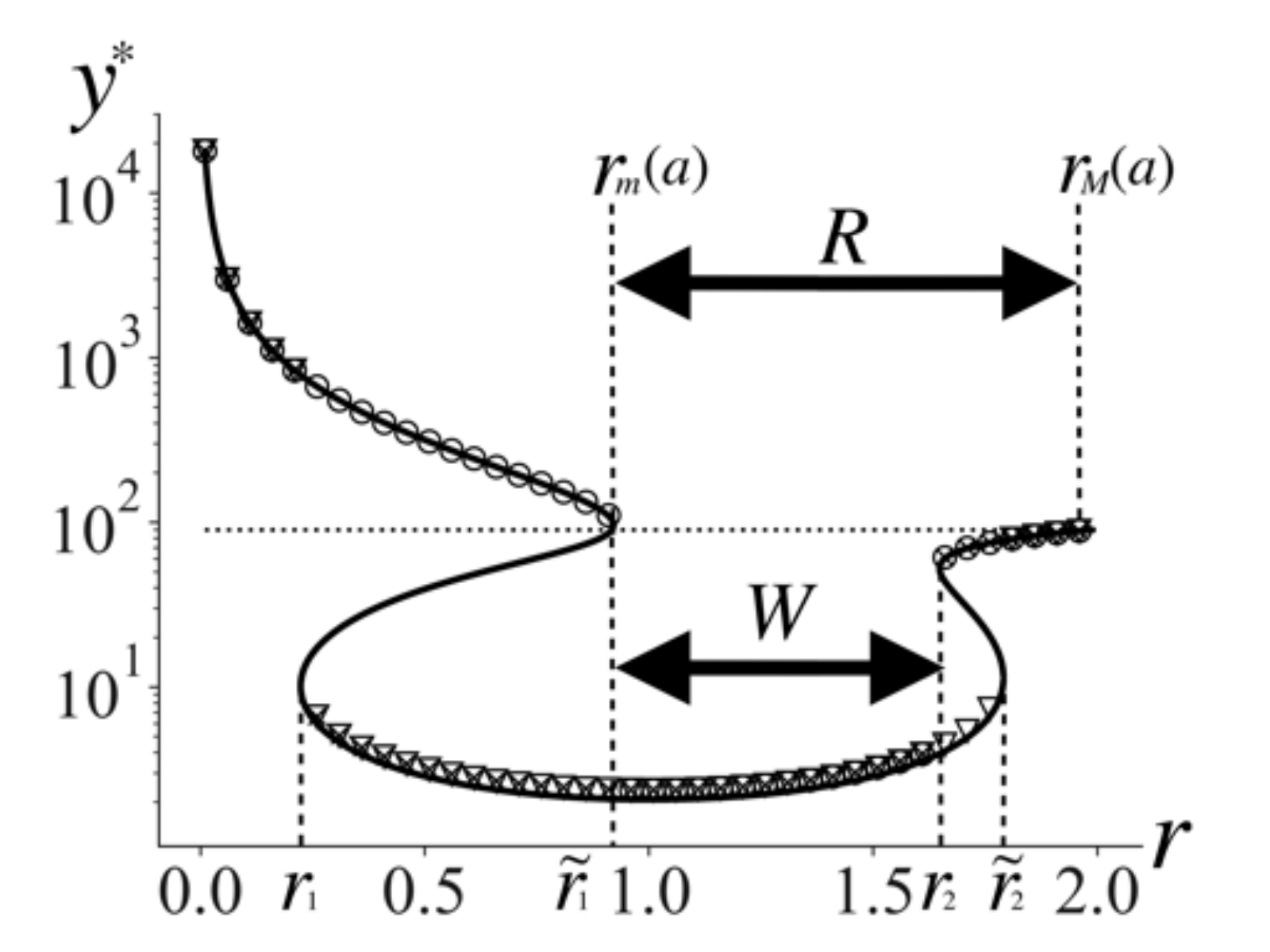}
\caption{Bifurcation diagram for the mutualistic ant--aphid system with facultative predation ($a=0.01$, $b=0.3$, $c=2.0$, $d=0.01$, $h=0.3$, $j=10$, and $k=90$). The dotted line is the carrying capacity $k$ of aphids and the circles (crosses and inverted triangles, respectively) denote the equilibrium population $y^*$ of aphids calculated by the numerical simulations from the initial state $(x_0, y_0)=(10, 1000)$ (from $(10,0.01)$ and $(400, 10)$, respectively). The solid curve was obtained by solving Eq.~(\ref{EqA15}).}
\label{Fig3}
\end{figure*}

We mathematically analyzed the system of Eqs.~(8) and (9) and obtained two trivial fixed points $(x,y)=(0,0)\equiv \vec{P}_0$ and $(x,y)=(0,k)\equiv \vec{P}_a$, and one or three internal fixed points $\vec{P}_I=(x^*>0,y^*>0)$ when $c-r>a$. Here, the homing rate $a$ is smaller than the balance $c-r$ of the resource for the honeydew reward for ants under the assumption that $b$ is small enough, that is, the encounter rate of ants $e_{yx}$ and the average handling time by ants $t_h$ are both large enough. 

We obtained the local stability condition for the trivial fixed points and we found that $\vec{P}_0$ is a saddle point for any positive values of $a$ and $r$, and $\vec{P}_a$ is locally stable when $c-r<a$, that is, when the system has no internal fixed point.
The former result means that the ants do not come to the host plant when it has no aphids and that the aphids grow independently if there are no ants initially. On the other hand, the latter result means that if the homing rate $a$ of ants is large enough and the balance $c-r$ (the resource distribution for the honeydew reward for ants) is small enough, the system converges to $\vec{P}_a$, that is no ants and $k$ aphids. Details of these analyses are given in \ref{AppendixA1}.

Assuming that $b$ is sufficiently smaller than $y$, we obtained the internal fixed point $\vec{P}_I=(x^*, y^*)$ and proved that the system has one or three internal fixed points. Details of these analyses are given in \ref{AppendixA2}. 

Since the intrinsic growth rate $r$ of aphids is the most important parameter for  the qualitative behavior of the system, 
we show the flows, fixed points, and $x$- and $y$-nullclines in the $xy$ phase space in Fig.~\ref{Fig2}, 
and the bifurcation diagram for $r$ in Fig.~\ref{Fig3}.
In Figs.~\ref{Fig2}(a)--(h) and Fig.~\ref{Fig3}, we observe that the system has two saddle node bifurcations at the first (second) bifurcation point $r=r_1\simeq 0.23$ ($r_2\simeq1.65$), and two inverse bifurcations at the first (second) inverse bifurcation point at $r=\tilde{r}_1\simeq 0.91$ ($\tilde{r}_2\simeq 1.79$), respectively.
Such bifurcations and the inverse ones are due to the cubic equation Eq.~(\ref{EqA15}), which gives rise to one or three internal fixed points, that is, they are due to the Holling's type III functional response $y^2/(j^2+y^2)$ for the facultative predation in Eq.~(\ref{Eq9}).

Note that even for a comparatively small value of $r=0.2$, the aphid equilibrium population reaches mutualistic coexistence 
($x^*\simeq 179, y^*\simeq 880$), which greatly exceeds the carrying capacity $k=90$ in Fig.~\ref{Fig2}(a).
On increasing the value of $r$, the first bifurcation occurs and another fixed point ($x^*\simeq 176, y^*\simeq 10$) emerges at $r_1$ (Fig.~\ref{Fig2}(b)). In the interval $r_1 \lesssim r \lesssim \tilde{r}_1$, the system has three internal fixed points, two of which are locally stable whereas the third is unstable (Fig.~\ref{Fig2}(c)). For $r\gtrsim \tilde{r}_1$, two fixed points merge (Fig.~\ref{Fig2}(d)) and the system has only one internal fixed point in the interval $\tilde{r}_1\lesssim r\lesssim r_2$ (Fig.~\ref{Fig2}(e)). Since the aphid equilibrium population $y^*$ is significantly lower than $k$ here (Fig.~\ref{Fig3}), we call this interval the valley of the population. $W\equiv r_2-\tilde{r}_1$ is the width of the valley. 
We further observe the second bifurcation at $r\simeq r_2$ (Fig.~\ref{Fig2}(f)). The system again has three fixed points in the interval $r_2 \lesssim r \lesssim \tilde{r}_2$ (Fig.~\ref{Fig2}(g)) and finally, for $r\gtrsim \tilde{r}_2$ the system has one internal fixed point $y^*\simeq k$ (Fig.~\ref{Eq2}(h)). 

Note that the vertical axis of Fig.~\ref{Fig3} is logarithmic. Thus, the valley is an order of magnitude deep and the aphids almost go extinct for $\tilde{r}_1\lesssim r\lesssim r_2$, since the aphids have insufficient resources.
This deep valley in the population of aphids is the distinguishing characteristic of the model with facultative predation. It predicts two scenarios for the mutualism between ants and aphids: (1) aphids with small $r\lesssim \tilde{r}_1$ and many ants or (2) aphids with large $r\gtrsim r_2$ and few ants. This may be why we generally observe two lineages of aphids, one associated with ants and the other not.

\begin{figure*}
 \begin{minipage}[b]{0.32\linewidth}
  \centering
  \includegraphics[scale=0.28,bb=0 0 461 346]  {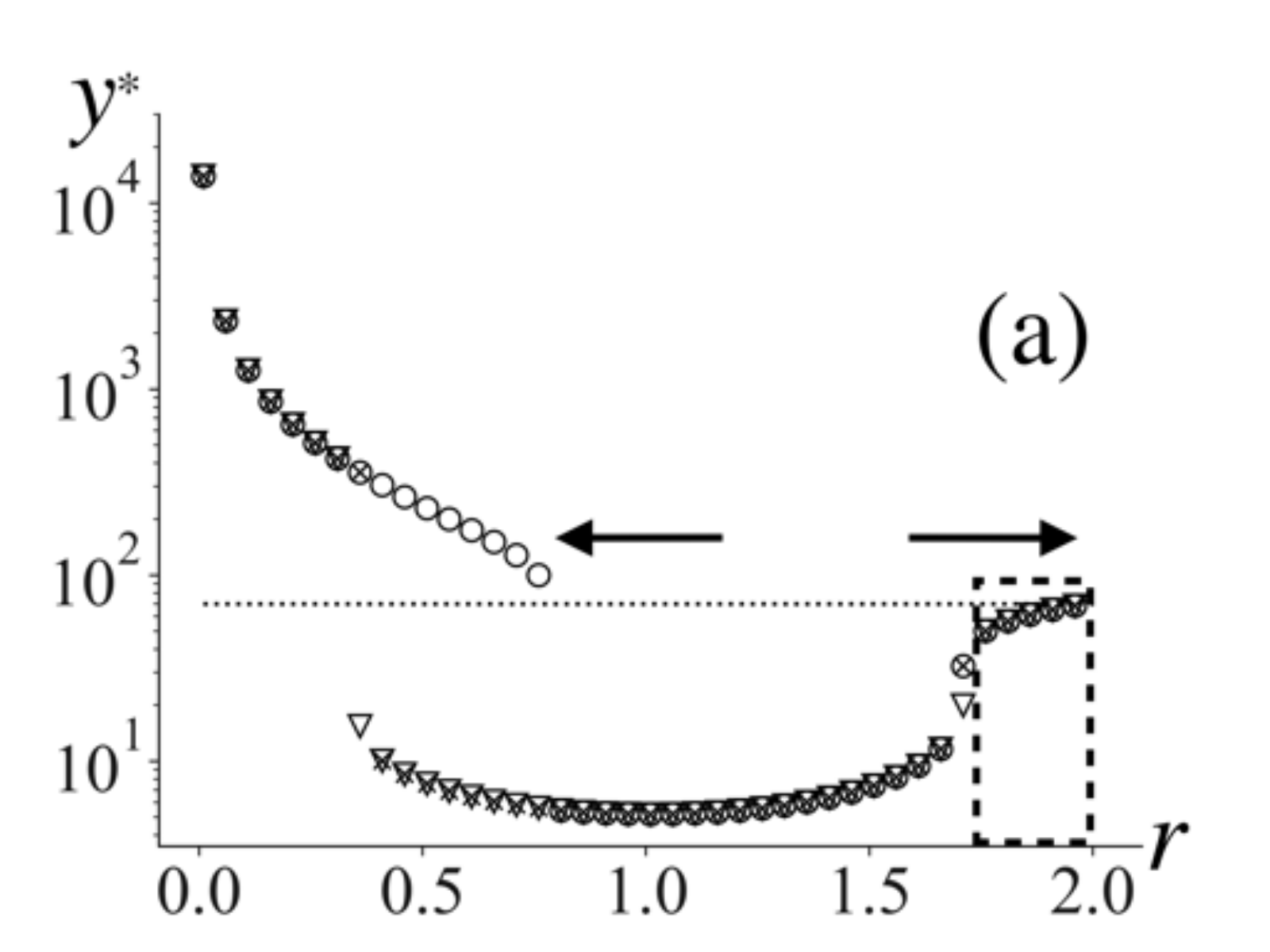}
 \end{minipage}
 \begin{minipage}[b]{0.32\linewidth}
  \centering
  \includegraphics[scale=0.28,bb=0 0 461 346]  {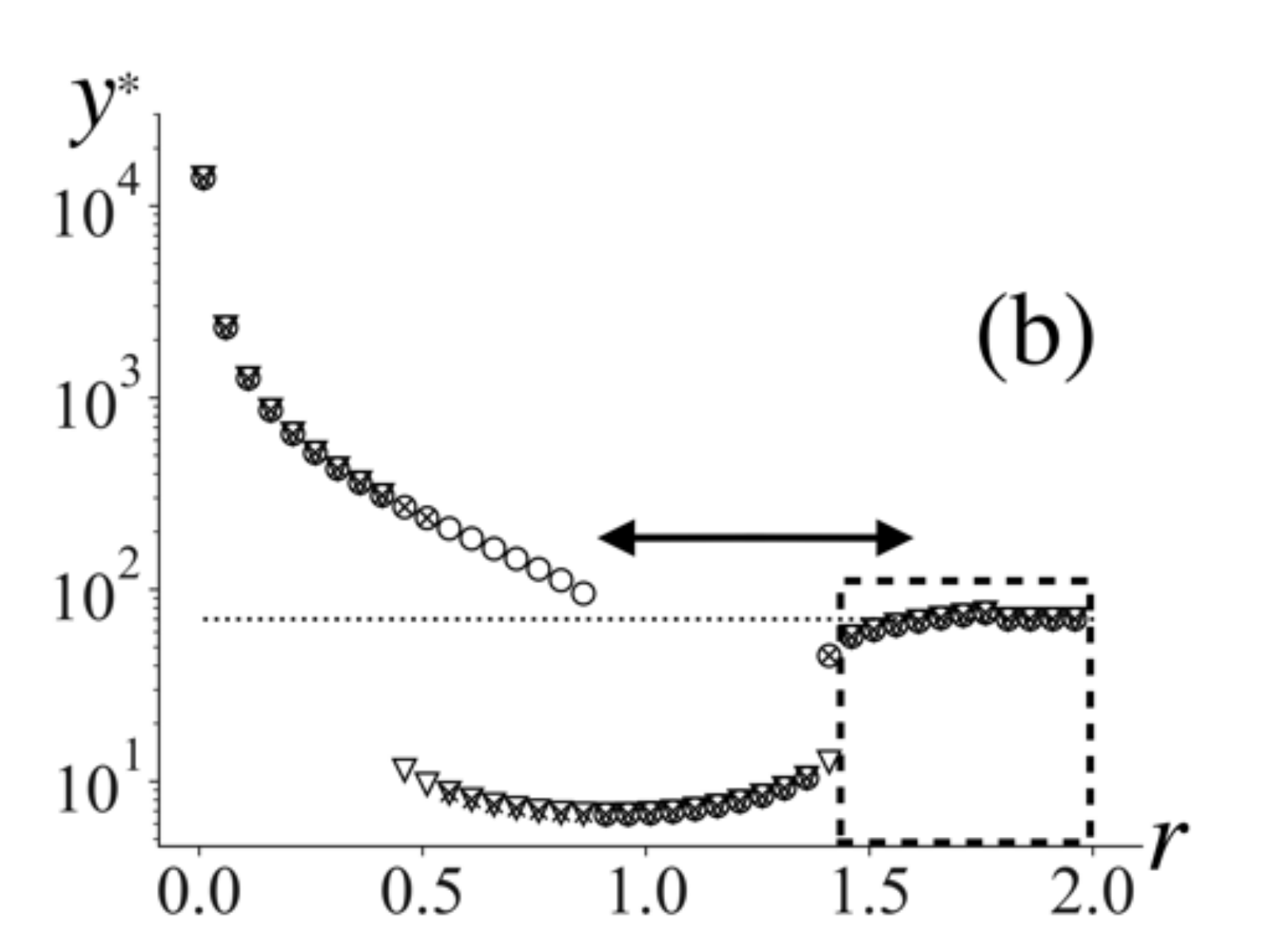}
 \end{minipage}
 \begin{minipage}[b]{0.32\linewidth}
  \centering
  \includegraphics[scale=0.28,bb=0 0 461 346]  {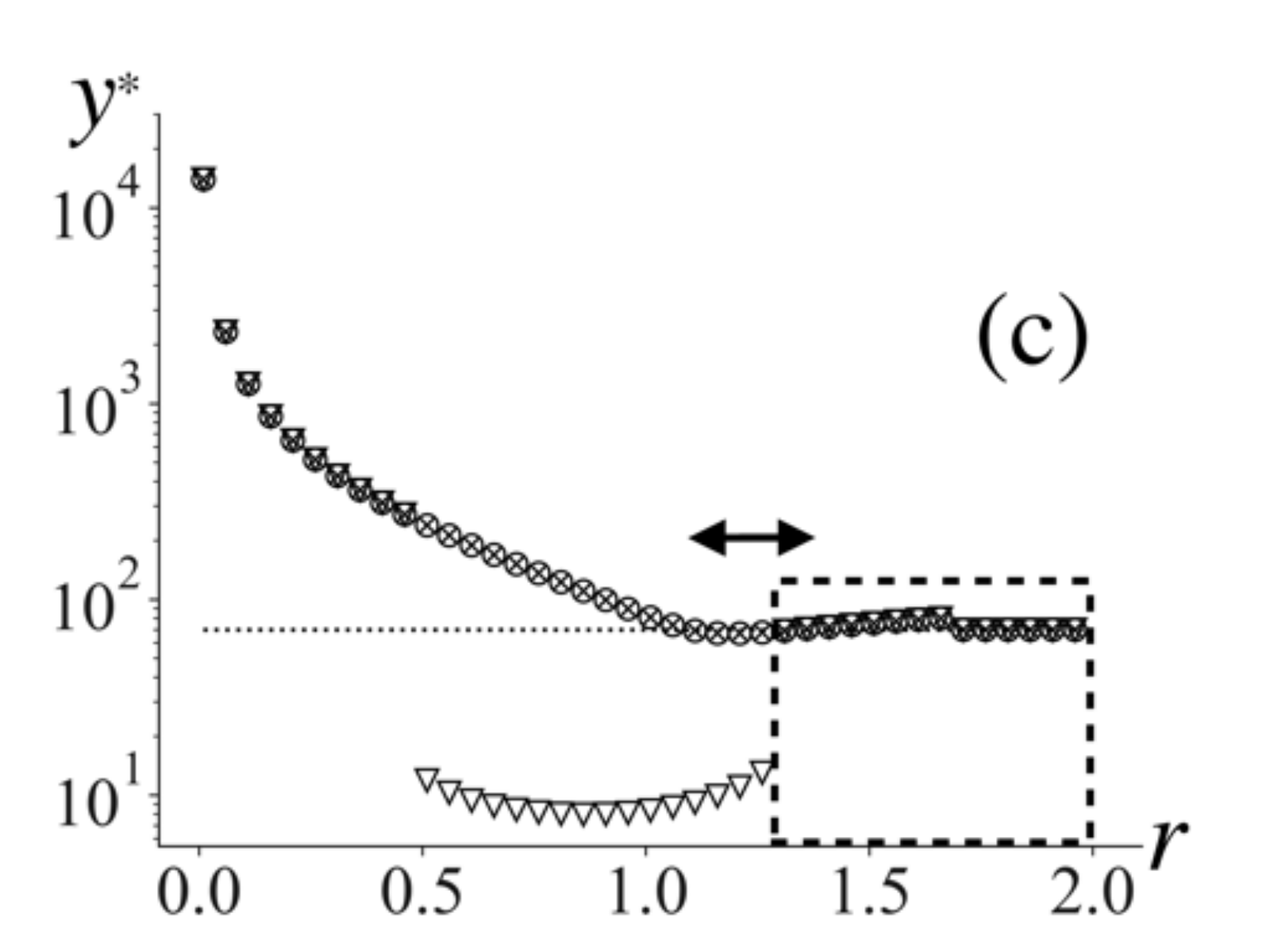}
 \end{minipage}
\caption{Bifurcation diagrams of mutualistic ant--aphid systems with facultative predation depending on the homing rate of ants: (a) $a=0.01$, (b) $a=0.2$, and (c) $a=0.3$. The values of the parameters are $b=0.3$, $c=2.0$, $d=0.01$, $h=0.3$, $j=15$, and $k=70$. The dotted rectangles highlight the range of $r$ where the equilibrium population $y^*$ of aphids is near the carrying capacity $k$.
The two-headed arrows in each panel denote the interval of $R$ obtained in \ref{AppendixB}.}
\label{Fig4}
\end{figure*}

Fig.~\ref{Fig4} shows the relation between the width of the valley and the homing rate $a$. 
As $a$ increases, the valley gets narrower. At $a = 0.3$, the valley disappears. Note also that the range highlighted by the dotted rectangle is longer for larger $a$. That is, for lower $r$, $y^*$ is higher and nearer to $k$. This means that if the ants have a higher homing rate, the hurdle for the aphids' ant-independence strategy is lower.
At first glance, this may seem to be a counterintuitive result, since a higher homing rate means abandoning the aphids, which may be expected to lead to a decline in the aphid population. 
However, this can be understood naturally by the facultative predation by the ants. 
If the ant homing rate increases and the number of attending ants decreases, then facultative predation, the third term on the right-hand side of Eq.~(\ref{Eq9}), weakens, and as a result, the aphid population is not in the bottom of the valley and it is near to the carrying capacity $k$. The mathematical analysis in \ref{AppendixB} indicates why the width of the valley is a function of $a$.

\begin{figure*}[p]
\centering
\includegraphics[scale=0.95,bb=0 0 432 360]{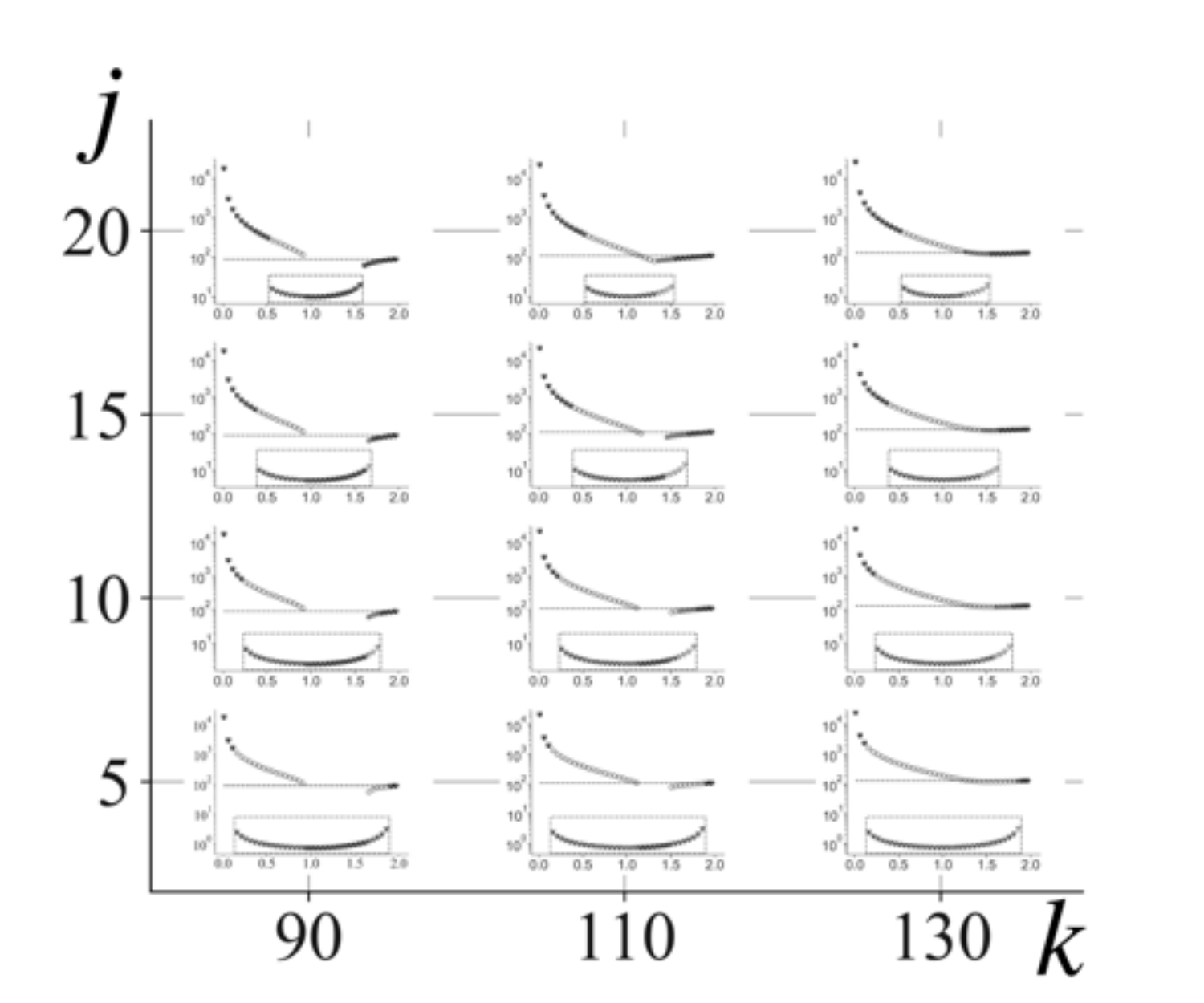}
\caption{Bifurcation diagrams for $j=5$, 10, 15, and 20, and $k=90$, 110, and 130. In each panel, the vertical axis is $y^*$ and the horizontal axis is $r$.
The values of the other parameters are $a=0.01$, $b=0.3$, $c=2.0$, $d=0.01$, and $h=0.3$.
The dotted rectangles highlight the range of $r$ where the equilibrium population of aphids $y^*$ is close to zero.}
\label{Fig5}
\end{figure*}

Fig.~\ref{Fig5} has bifurcation diagrams for $j=5$, 10, 15, and 20, and $k=90$, 110, and $130$. 
As the carrying capacity $k$ increases, the valley becomes narrower. 
In contrast, as the half-saturation population of aphids $j$ decreases, the range of $r$ for the lower branch of $y^*$ gets wider (the bottom of the valley or the endangered state), as highlighted by the dotted rectangles. 

We, moreover, analyzed the mutualistic ant--aphid system without facultative predation (Eqs.(\ref{EqB1})--(\ref{EqB2})). 
In this case, we obtained two trivial fixed points $(x,y)=(0,0)\equiv\vec{P}_0$ and $(x,y)=(0,k)\equiv\vec{P}_a$, and one internal fixed point $\vec{P}^{(n)}_I=(x^*>0, y^*>0)$ (Eq.~(\ref{EqB9})) when $c-r>a$, that is, the resource distribution $c-r$ for the honeydew reward for ants is larger than the homing rate $a$ of the ants. We also obtained the local stability condition of the trivial fixed points and found that $\vec{P}_0$ is a saddle and $\vec{P}_a$ is locally stable when $c-r<a$, which is the same as for the system with facultative predation. Details of the analyses are given in \ref{AppendixC1}. If $b$ is sufficiently smaller than $y$, then $\vec{P}_I^{(n)}$ is locally stable when $c-r>a$. The details are also given in \ref{AppendixC2}. In addition, we proved that the model without facultative predation 
has no closed orbit in the positive quadrant $x, y>0$ (\ref{AppendixC3}).

\begin{figure*}[h]
 \begin{minipage}[b]{0.5\linewidth}
  \centering
  \includegraphics[scale=0.37,bb=0 0 461 346]
  {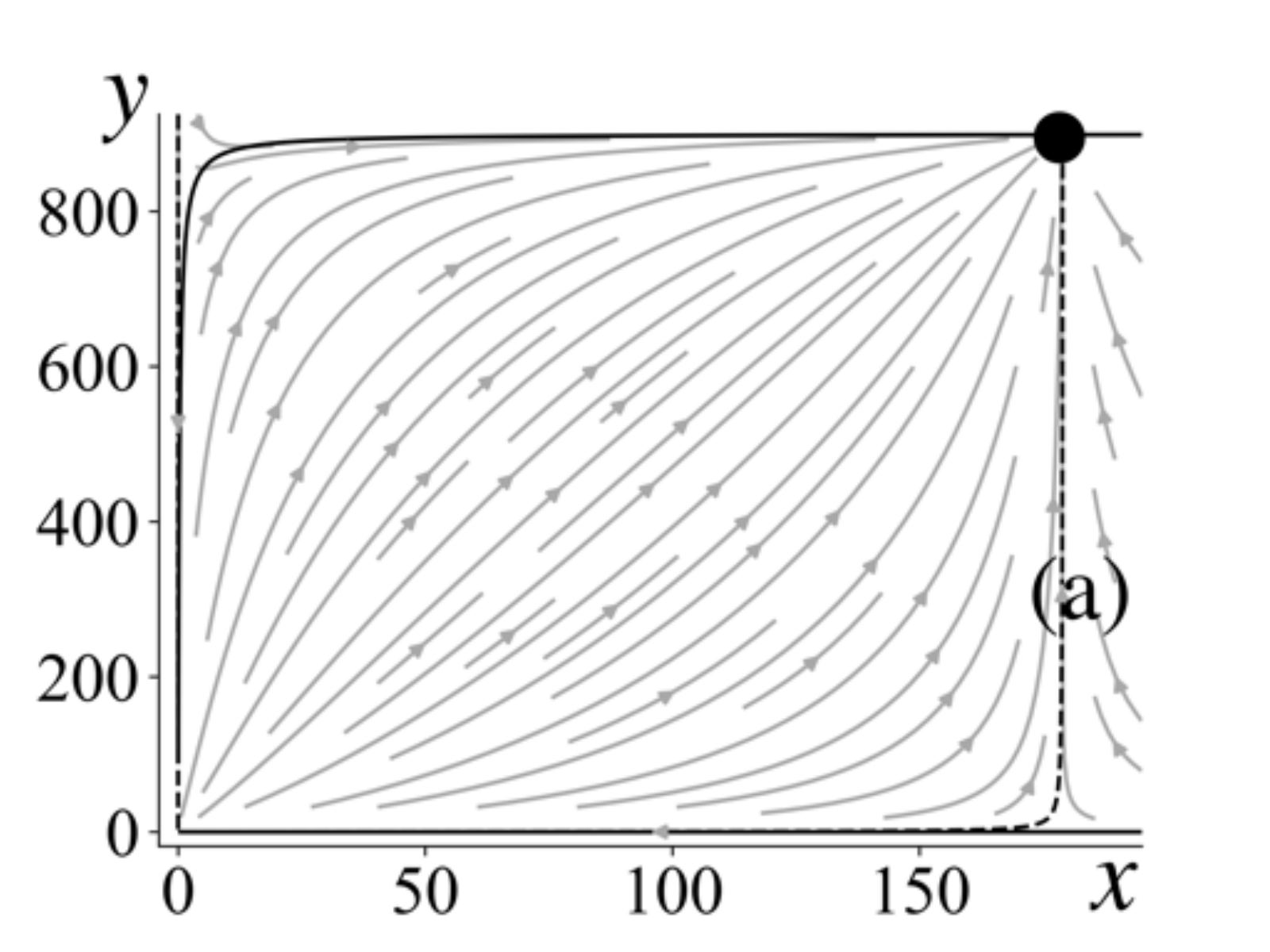}
 \end{minipage}
 \begin{minipage}[b]{0.5\linewidth}
  \centering
  \includegraphics[scale=0.37,bb=0 0 461 346]
  {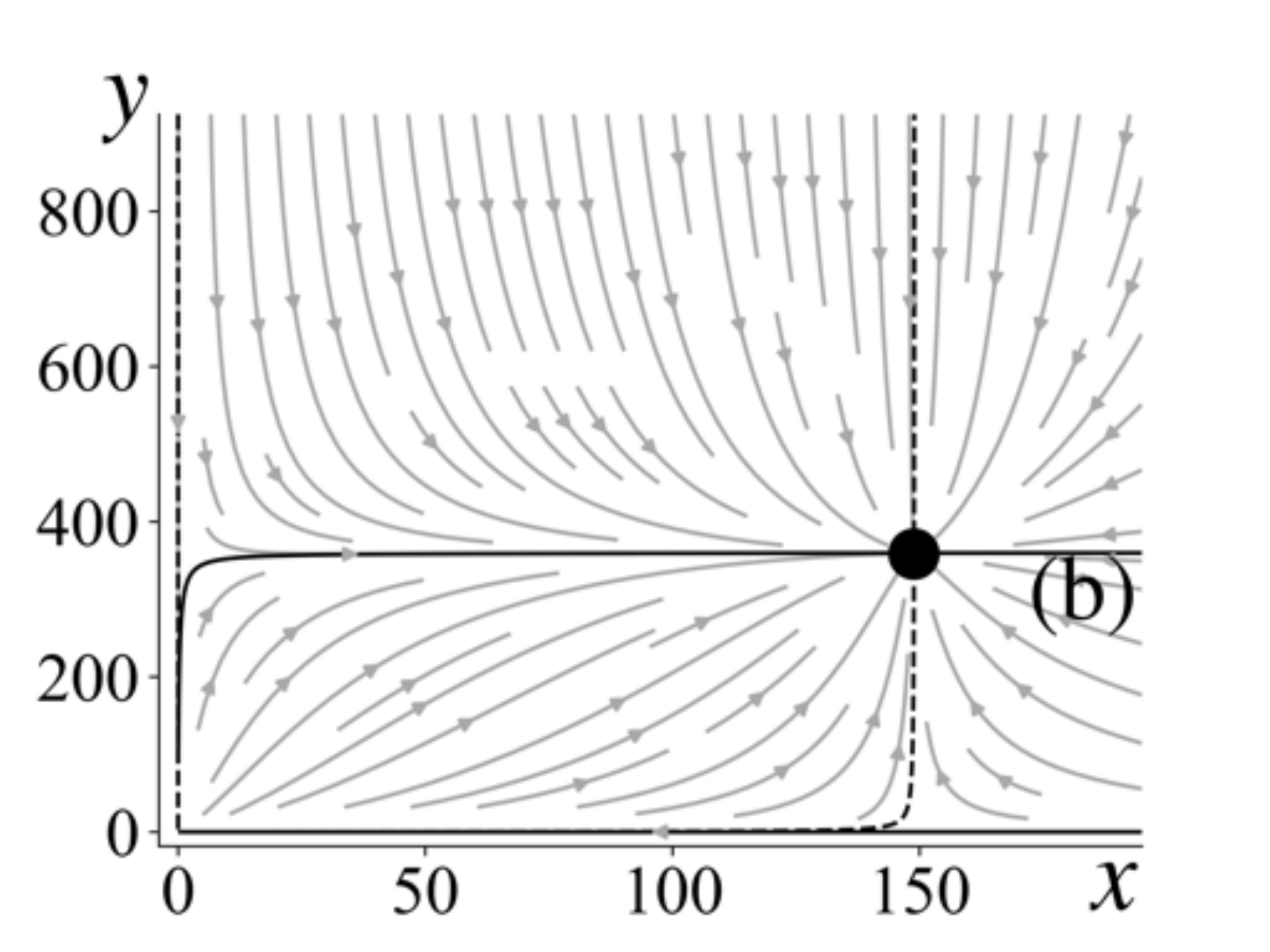}
 \end{minipage}\\
 \begin{minipage}[b]{0.5\linewidth}
  \centering
  \includegraphics[scale=0.37,bb=0 0 461 346]
  {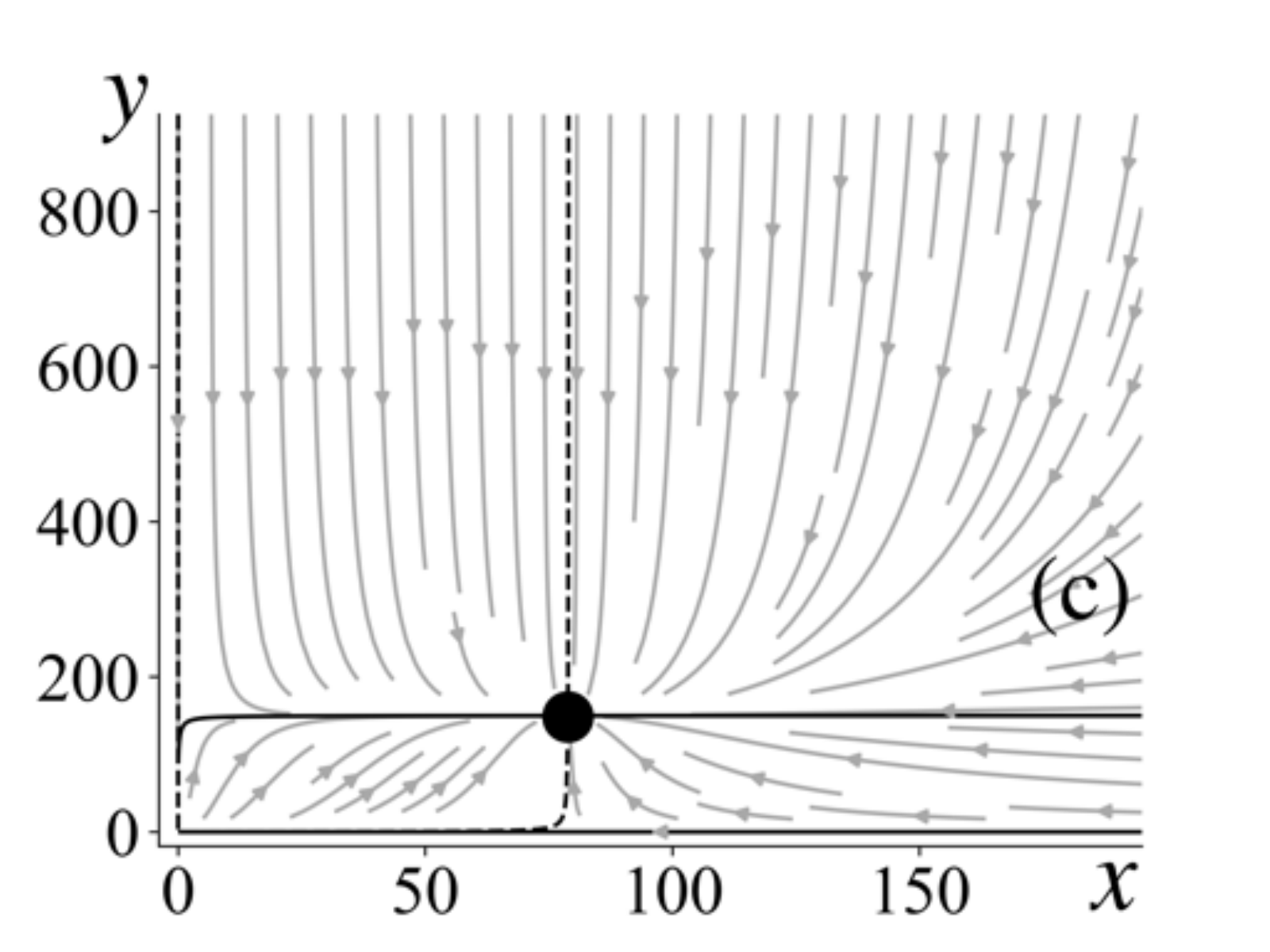}
 \end{minipage}
 \begin{minipage}[b]{0.5\linewidth}
  \centering
  \includegraphics[scale=0.37,bb=0 0 461 346]
  {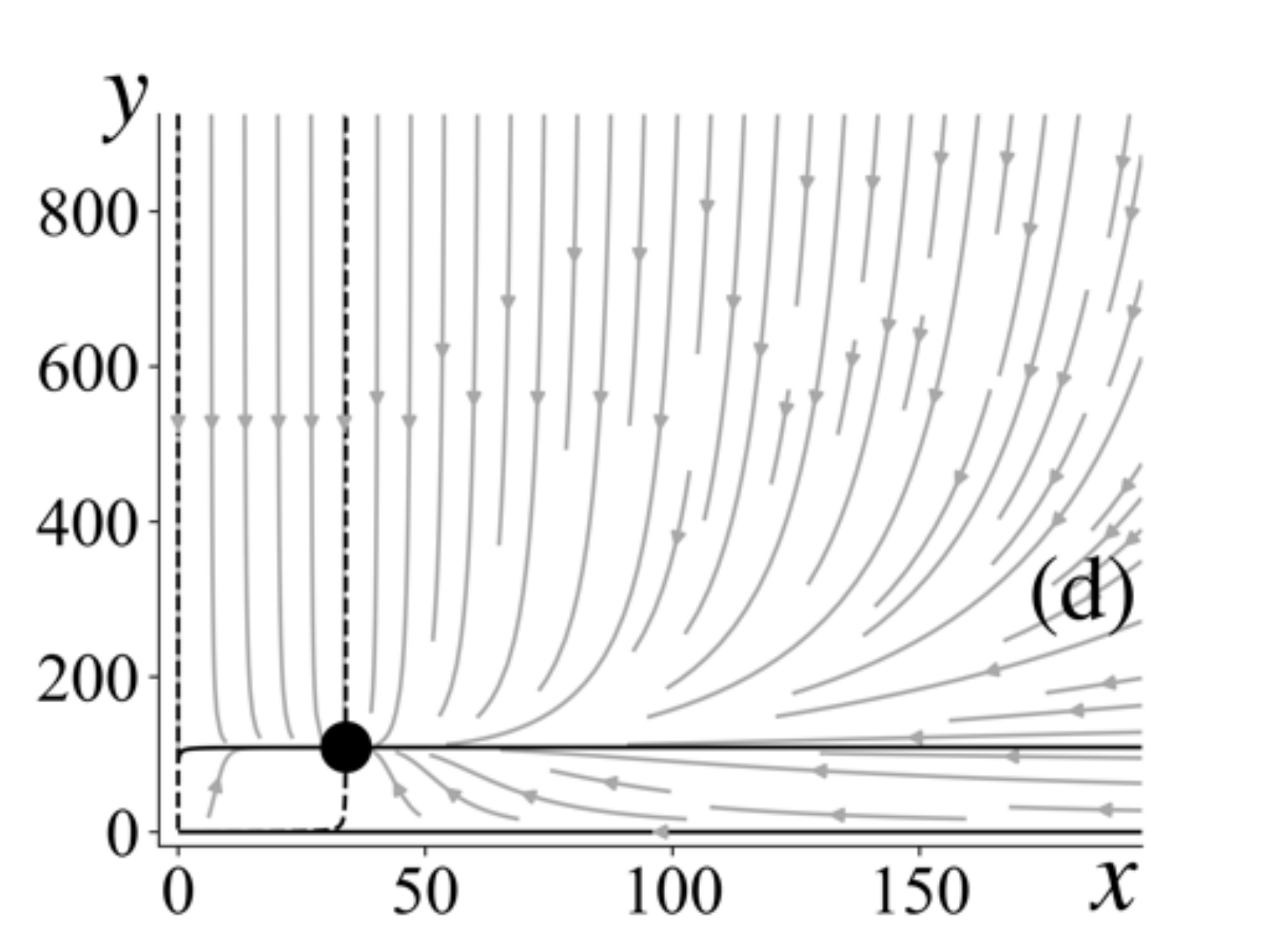}
 \end{minipage}\\
 \begin{minipage}[b]{0.5\linewidth}
  \centering
  \includegraphics[ scale=0.37,bb=0 0 461 346]
  {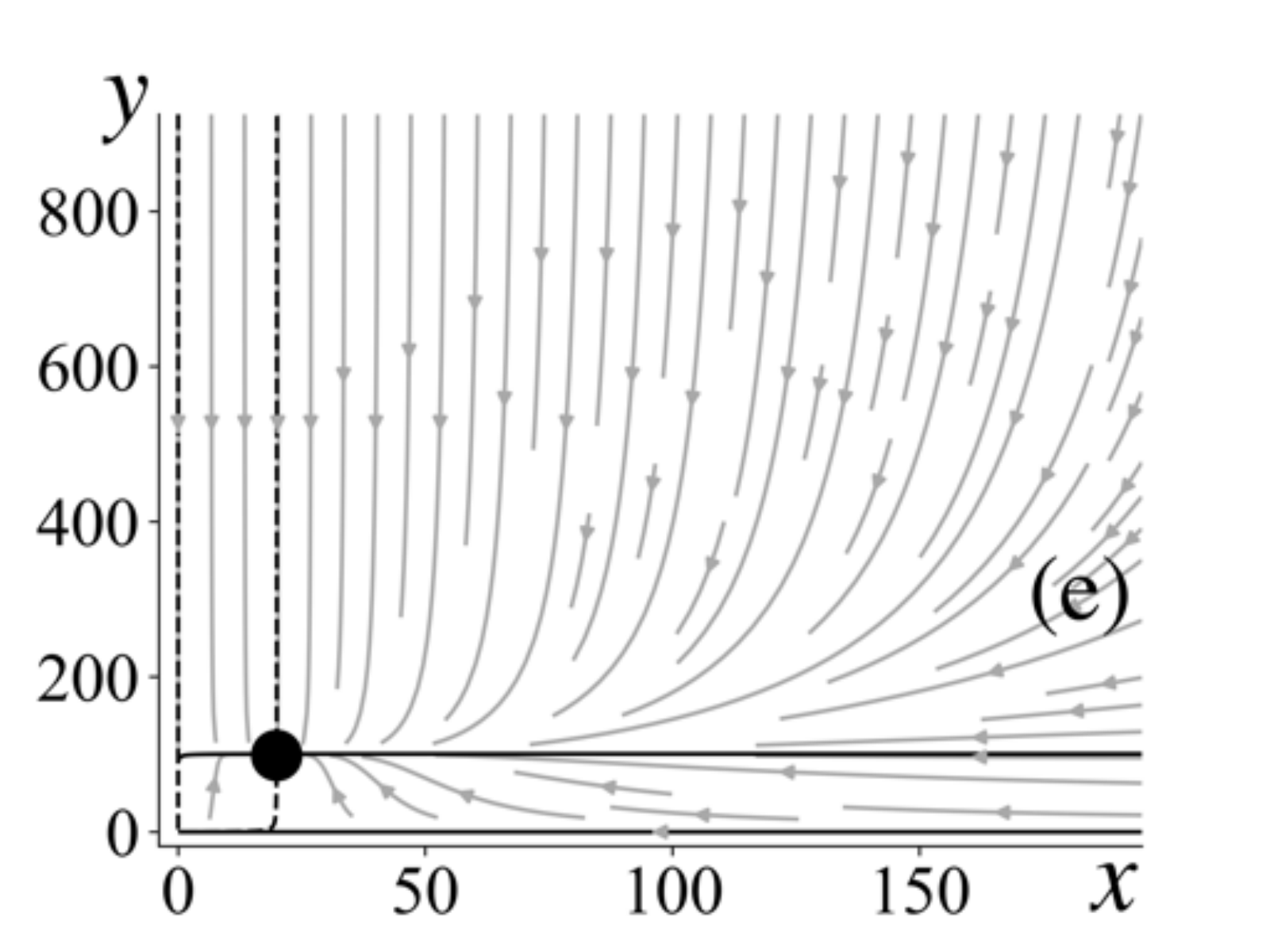}
 \end{minipage}
  \begin{minipage}[b]{0.5\linewidth}
  \centering
  \includegraphics[scale=0.37, bb=0 0 400 346]
  {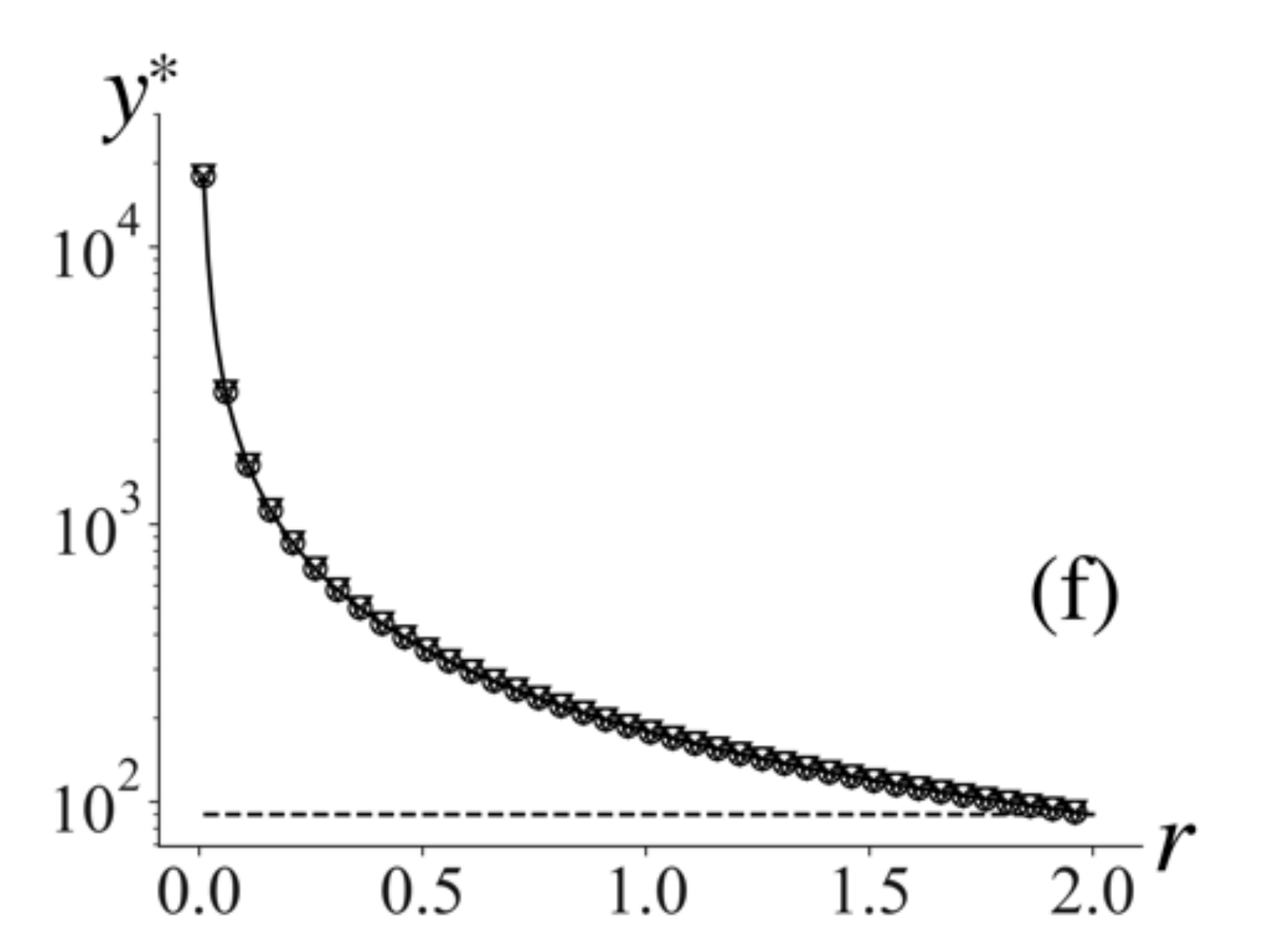}
 \end{minipage}
 \caption{(a)--(e) Flows, with locally stable fixed points (black circles), $x$-nullclines (dashed lines), $y$-nullclines (solid lines), and (f) bifurcation diagram for the mutualistic ant--aphid system without facultative predation. The values of the parameters are $a=0.01$, $b=0.3$, $c=2.0$, $d=0.01$, $h=0.3$, and $k=90$, and (a) $r=0.2$, (b) $r=0.5$, (c) $r=1.2$, (d) $r=1.65$, and (e) $r=1.79$. In (f), the dotted line shows the carrying capacity $k$ of aphids and the circles (crosses and inverted triangles, respectively) denote the equilibrium population $y^*$ of aphids calculated by the numerical simulations from the initial state $(x_0, y_0)=(10, 1000)$ (from $(10,0.01)$ and $(400, 10)$, respectively). The solid curve in (f) is $y^*$ from Eq.~(\ref{EqB8}).}
\label{Fig6}
\end{figure*}

Fig.~\ref{Fig6} shows the flows in the $xy$ phase space, the $x$- and $y$-nullclines, and the graph of $y^*$ as a function of $r$ for the system without facultative predation. In contrast to the system with facultative predation (Figs.~\ref{Fig2} and \ref{Fig3}),  in Fig.~\ref{Fig6}(f) we observe neither a bifurcation nor the valley of the population, which means that as $r$ gets smaller, $y^*$ gets larger.
Thus, there is a simple win--win-type relation between ants and aphids. That is, the population of aphids is more abundant if they allocate more resources $c-r$ to the honeydew reward for ants than to their own reproduction~$r$.

\section{Discussion}
Here, we review the novelty of the modeling and the results of the present study, in comparison with a previous mathematical model \citep{Holland2010}. There are three essential differences between the previous models and ours. The first point is the difference in the scope of the modeling. The previous authors studied the general mathematical properties of a mix of mutualistic and predatory relationships without considering any particular species, whereas our model is based solely on the ant--aphid system. 
The second point is that they assumed the Holling's type II functional response for predation whereas we assumed the type III functional response. 
The third point is the most important and makes a decisive difference in prediction. That is, there is a trade-off between the parameters for mutualism and facultative predation in our model whereas the parameters for mutualism and predation were independent of each other in the consumer--resource interaction model. Based on this difference, we observe comparatively complicated bifurcations and the valley of the population, which was not found in the consumer--resource interaction model.

In the context of mathematical modeling of mutualism with facultative predation, there is a difference in the results when assuming the Holling's type II functional response instead of the type III functional response for predation. In fact, we did preliminary research on such a model and found more complicated flows, including stable spirals and more complicated bifurcations. These need a more elaborate analysis because we cannot use the approximation that the value of $b$ is sufficiently small. We, nevertheless, observed the valley of the population, too, in such a model, which suggests that the main results here are robust to the variation of the functional form of facultative predation. In modeling the ant--aphid system, the type III functional response is more appropriate for facultative predation than the type II functional response, but there may be other mutualistic systems with facultative predation that are best modeled by the type II functional response. Theoretically, such a mathematical structure based on the type II functional response requires further detailed analysis.

The facultative predation of aphids by ants is, also ecologically, inferred to be Holling's type {III}.
Ants that attend aphids and harvest honeydew largely belong to the subfamilies Formicinae and Dolichoderinae, plus a few genera of Myrmicinae \citep{holldobler1990ants,nixon1951association}. Many of these ants are generally categorized as predators and scavengers and they collect a wide variety of arthropods as food for their colony \citep{carroll1973ecology,mooney2005temporal}. The ants may be relatively less dependent on attending aphids as a nitrogen source although they largely depend on honeydew as a sugar source. In addition, ants tend to take prey insects to their colony instead of immediately consuming and digesting them. These feeding habits of the aphid-attending ants also led us to assume the type III functional response for predation by ants. 

One of the non-trivial theoretical predictions of this study is that as the ant homing rate increased, the hurdle for the aphids' ant-independent strategy decreased (Fig.~\ref{Fig4}). In other words, when ants begin depending on other aphids or reduce their dependence on honeydew from aphids, the homing rate increases, and in this situation, the aphid population increases even though the value of the intrinsic growth rate of aphids and the resource allocation for the honeydew reward for ants are both unchanged. It is as if the aphids do not expect single-minded ants. Is this paradox of mutualism simply due to modeling failure, or is it actually possible in nature? Again, the concept of facultative predation provides an evolutionary ecological answer to this question. The answer is that mutualism evolved after predation had evolved for the ant--aphid system. Originally, ants were simply a predator of aphids, they preyed on aphids arbitrarily, their homing rates were high (Fig.~\ref{Fig4}(c)), and the aphids did not allocate any resource to a honeydew reward for the ants ($r\sim c$). Subsequently, a new lineage of aphids that depended on ants for defense emerged. These aphids increased the resource allocated to the  honeydew reward for ants, $c-r$, and the intrinsic growth rate of aphids $r$ started to decline in this ant--aphid system. As this specialization of the relationship between ants and aphids advanced, the decline of $r$ continued while the homing rate was high, and highly ant-dependent aphids prospered above the carrying capacity $k$ (Fig.~\ref{Fig4}(c); $r\lesssim 1.0$). At the same time, this specialization reduced the homing rate of ants. Typical examples of such specialized ant--aphid systems involve obligate myrmecophilous species of {\it{Stomaphis}} aphids, which are harbored in shelter-like trails of several species of {\it {Lasius}} ants \citep{blackman1994aphids,takada2008life}. {\it {Lasius}} ant workers routinely care for the aphids and harvest honeydew, so that their homing rate is quite low. With the lowering of the homing rate, the valley of the population could form (Figs.~\ref{Fig4}(a) and~\ref{Fig4}(b)) due to the trade-off between the intrinsic growth rate of aphids and the honeydew reward for ants, and due to facultative predation. At the same time, the extinction rates of aphids with intermediate values of $r$ increased, and ant-dependent and ant-independent aphids differentiated. 

To summarize, the scenario in which the hurdle of aphids' ant-independent strategy lowers as the ants' homing rate increases can be considered as a rewind of the above evolutionary history. This scenario of mutualism arising due to facultative predation could be demonstrated by a future experimental study qualitatively comparing the quality of honeydew and the preference of ants for honeydew for two types of ant--aphid system: (1) a system with a weak relationship between the ants and aphids (e.g., a variety of ants that attends and uses a range of aphids) and (2) a system with strong relationship (e.g., a specific and one-to-one relationship between ants and aphids). Furthermore, if the relevant genes can be specified, a phylogenetic analysis is then possible.

In a system without facultative predation by the ants, both populations can, in theory, grow better at smaller intrinsic growth rate $r$ of aphids (Fig. 6). Although we assumed that the ant population $x$ was restricted to a host plant, so that it does not reflect the whole colony, a smaller $r$ means the  aphids need to make a greater investment in producing attractive honeydew, which is nutritionally beneficial for the whole colony of the ants.
Thus, an ant population seems to have successfully developed without facultative predation on the aphids being attended. 
However, such a win--win outcome is unlikely for aphids, as they are an exceptionally highly $r$ insect group. In addition, facultative predation is likely to prevail among aphid-attending ants because aphids are primarily a common prey species, even for aphid-attending ants \citep{skinner1980feeding,mooney2005temporal} and ants are usually a major insect predator. The results from our model with predation suggest that facultative predation by aphid-attending ants has an important role in ant--aphid population dynamics, which has been overlooked in previous studies \citep[e.g.,][]{Holland2010}. Facultative predation means the aphids adopt either of roughly two strategies: being dependent on mutualism or being dependent on high $r$.

In the context of evolutionary theory, one limitation of the present study is that we considered only population dynamics with fixed values for the ecological parameters and obtained results predicting that  two lineages of aphids can thrive and that aphids with an intermediate $r$ have the highest extinction rate. To clarify whether two such lineages can be actually differentiated from one another, we need to investigate adaptive dynamics \citep{dieckmann1996dynamical,geritz1998evolutionarily,otto2007}, evolutionary dynamics with many lineages \citep{nowak1991mathematical}, or evolutionary dynamics on the space of genetic traits \citep{sasaki1994evolution}. The modeling of such adaptive dynamics or evolutionary dynamics of an ant--aphid system with many lineages would be more elaborate and need more parameters but is a promising challenge for the future.

Finally, extending the model to include host plants as well as aphids and ants would be promising future work because there has been interesting work on the interactions between aphids and host plants, and on the maintenance of aphid polymorphism \citep{watanabe2016color,watanabe2018effects,watanabe2018ants}. 

\section{Conclusions}

We have demonstrated that an ant--aphid population qualitatively depends on facultative predation by ants and by the trade-off for aphids between allocating their resources between the intrinsic growth rate and secreting a honeydew reward for the ants. The main conclusions and theoretical predictions of this paper are summarized as follows. A moderate dependence on ants may increase the aphid extinction rate. Aphids do not require single-minded ant attendance. Two lineages of aphids -- those being attended by ants or not -- can evolve. The facultative predation of aphids by ants may be an example of a Holling's type III functional response in nature. These insights are expected to result in a new understanding of mutualism. Future experimental studies are required to verify them.

\section*{Authors' contributions}
AN, YI, and KT conceived the idea. AN and YI performed the literature search and analyzed the data.
AN and KT contributed to the mathematical analysis, visualization, and interpretation of the results.
AN wrote the original draft, and YI and KT contributed to reviewing and editing the manuscript.

\section*{Declaration of Competing Interests}
The authors declare no conflicts of interest.

\section*{Acknowledgments}
In launching and promoting this research, discussions with the following people were illuminating. The authors are deeply grateful to Prof. E. Hasegawa (Hokkaido University), Prof. T. Namba (Osaka Prefecture University), Prof. J. Yoshimura (Shizuoka University), Dr. S. Watanabe (Kyoto University), and Dr. Y. Uchiumi (SOKENDAI). This work was partially supported by the Research Institute for Mathematical Sciences, International Joint Usage/Research Center, Kyoto University, and KAKENHI through grants 16K07516 (YI) and 19K03650 (KT).
The authors thank the anonymous reviewers for their constructive comments, which helped to improve the manuscript significantly.

\appendix
\section{Fixed-point analysis for the model with facultative predation}
The elements $a_{ij}$ of the Jacobian $A$ of the system (\ref{Eq8}) and (\ref{Eq9}) are 
 \begin{align}
 a_{11}&=- a - 2 d \cdot x +  \frac{y \left(c - r\right)}{b + y}, \\
 a_{12}&=-  \frac{x y \left(c - r\right)}{\left(b + y\right)^{2}} + \frac{x \left(c - r\right)}{b + y}, \\
 a_{21}&=- \frac{r y^{2}}{j^{2} + y^{2}} - \frac{x y \left(c - r\right)}{\left(h + x\right)^{2}} + \frac{y \left(c - r\right)}{h + x},\\
 a_{22}&= \frac{2 r x y^{3}}{\left(j^{2} + y^{2}\right)^{2}} - \frac{2 r x y}{j^{2} + y^{2}} + r \left(1 - \frac{y}{k}\right) + \frac{x \left(c - r\right)}{h + x} - \frac{r y}{k}.
 \end{align}
We analyzed the stability of the fixed points using these equations.
 
\subsection{Local stability of the trivial fixed points}\label{AppendixA1}
At one of the trivial fixed points $\vec{P}_0=(x,y)=(0,0)$,
the Jacobian $A_{(0,0)}$ is given by
\begin{align}
A_{(0,0)}=\left[\begin{matrix}- a & 0\\0 & r\end{matrix}\right].
\end{align}
Since its determinant is $\operatorname{det}A_{(0,0)} =-ar<0$, $\vec{P}_0$ is a saddle point. 
This means that ants do not come to the host plant when there are no aphids and that the aphids grow independently if there are no ants initially.  

The Jacobian $A_{(0,k)}$ of 
another trivial fixed point $\vec{P}_a=(x,y)=(0,k)$ is given by
\begin{align}
A_{(0,k)}=\left[\begin{matrix}- a + \dfrac{k \left(c - r\right)}{b + k} & 0\\- \dfrac{k^{2} r}{j^{2} + k^{2}} + \dfrac{k \left(c - r\right)}{h} & - r\end{matrix}\right],
\end{align}
and the determinant and the trace are
\begin{align}
\operatorname{det}A_{(0,k)}&=- r \left(- a + \frac{k \left(c - r\right)}{b + k}\right),\\
\operatorname{Tr}A_{(0,k)}&=- a + \frac{k \left(c - r\right)}{b + k} - r.
\end{align}
The local stability conditions for $\vec{P}_a$ are given by
$\operatorname{det}A_{(0,k)}>0$ and $\operatorname{Tr}A_{(0,k)}<0$. 
Thus, $\vec{P}_a$ is locally stable when $k(c-r)/(b+k)<a$ and $k(c-r)/(b+k)<a+r$, that is,
 \begin{align}
\frac{k(c-r)}{b+k}<a.
\end{align}
Assuming that $b$ is sufficiently smaller than $k$, the above condition can simply be approximated as 
\begin{align}
c-r<a . \label{EqA10}
\end{align}
Therefore, $\vec{P}_a$ is a locally stable fixed point when (\ref{EqA10}) holds.

\subsection{Internal fixed points}\label{AppendixA2}
From Eq.~(\ref{Eq8}), the equations for the $x$-nullclines are 
\begin{align}
x&=0,\\ 
x&=-\frac{a}{d}+\left(\frac{c-r}{d}\right) \frac{y}{b+y}. \label{EqA.12}
\end{align} 
Assuming that $b$ is sufficiently smaller than $y$, Eq.~(\ref{EqA.12}) can be rewritten as
\begin{align}
x=\frac{c-r-a}{d}. \label{EqA13}
\end{align}
If an internal fixed point $\vec{P}_I=(x,y)=(x^*>0,y^*>0)$ exists, from Eq.~(\ref{EqA13}) the condition $c-r>a$ should hold.
That is, when the homing rate $a$ is smaller than the balance $c-r$ of the resource for the honeydew reward for ants, there is a solution for which a non-zero number of ants attend aphids and $\vec{P}_a$ is unstable. On the other hand, when the homing rate $a$ and the intrinsic growth rate $r$ of aphids are both high enough ($c-r<a$), there is no internal fixed point and $\vec{P}_a$ is stable, that is, all ants return to their nest and the aphids grow by themselves.

The equation for the $y$-nullcline is 
\begin{align}
r\left(1-\frac{y}{k}\right)+(c-r)\frac{x}{h+x}-r\left(\frac{y}{j^2+y^2}\right)x=0.\label{EqA14}
\end{align}
Substituting $x=(c-r-a)/d$ (\ref{EqA13}) into (\ref{EqA14}) and assuming that $b$ is effectively smaller than $y$, we obtain the following equation:
\begin{multline}
-\frac{r}{k}y^3 +\left\{r + \frac{(c-r)(c-r-a)}{h d + c - r - a}\right\} y^2
-\left(\frac{rj^2}{k} + r\frac{c-r-a}{d}\right) y \\
\qquad +\left\{rj^2+(c-r)j^2\frac{c-r-a}{h d + c-r -a}\right\}=0 .\label{EqA15}
\end{multline}
Solving this cubic equation yields $y$ but the solution is not included here because it is too long. 
The values of the internal fixed point $\vec{P}_I=(x^*, y^*)$ obtained from Eqs.~(\ref{EqA13}) and (\ref{EqA15}) match those obtained by directly simulating Eqs.~(\ref{Eq8}) and (\ref{Eq9}) with high precision, which can be confirmed by Fig.~\ref{Fig3}.

Consequently, by considering that all parameters are positive, $c-r>a$ is the existence condition for $\vec{P}_I$. The signs of each polynomial coefficient for powers of $y$ in (\ref{EqA15}) are 
\begin{align}
0 &>-\frac{r}{k},\\
0 &<r+\frac{(c-r)(c-r-a)}{h d +c -r -a},\\
0 &>-\left(\frac{rj^2}{k} + r \frac{c-r-a}{d}\right),\\
0 &<rj^2+(c-r)j^2 \frac{c-r-a}{h d +c-r-a}.
\end{align}
There are three sign changes: (1) $-\to +$, (2) $+\to -$, and (3) $-\to +$.
Therefore, by Descartes's rule of signs, we conclude that the number of positive real solutions (including multiple solutions) of Eq.~(\ref{EqA15}) is three or one, that is
the system (\ref{Eq8}) and (\ref{Eq9}) has one or three internal fixed points depending on the parameters.

\section{Bifurcation points and width of the valley of the population}\label{AppendixB}
Since the equation for the bifurcation points is not an algebraic equation of the fourth or lower order and it is impossible to obtain the bifurcation points analytically, we calculate them approximately here and estimate the width of the valley of the population.
From Fig.~\ref{Fig3}, we observe that the left rim of the valley (the first inverse bifurcation point $\tilde{r}_1$) is at $(r,y^*)\simeq (\tilde{r}_1, k)$ 
and the value of $y^*$ at the right rim (the second bifurcation point $r_2$) is slightly less than $k$. Thus, by finding the $r$ at which $y^*=k$, we can estimate the approximate value of the (inverse) bifurcation points and the width of the valley.

Using the approximation $b\ll y$, consider the condition of $r$ for a fixed point:
\begin{equation}
(x^*,y^*) \simeq \left(\frac{c-r-a}{d},k\right). \label{EqB.1}
\end{equation}
By inserting Eq.~(\ref{EqB.1}) into Eq.~(\ref{EqA14}), the equation for $r$ is 
\begin{align}
(c-r)\frac{d}{h d +c - r - a} - r\left(\frac{k}{j^2+k^2}\right)=0, \label{EqB.2}
\end{align}
where we used $x^*=(c-r-a)/d\neq 0$.
Let $A\equiv k/(j^2+k^2)>0$ and $B(a)\equiv h d+c-a$, and considering only the case $B(a)>0$ (fulfilled for the parameter sets used in the present study, e.g., in Figs.~\ref{Fig2}--\ref{Fig4}), we obtain a quadratic equation in $r$:
\begin{align}
Ar^2-K(a)r+L&=0,  \label{EqB.5}
\end{align}
where $K(a)\equiv AB(a)+d>0$ and $L\equiv cd>0$.
By solving this and if $K(a)^2-4AL>0$  (fulfilled in Figs.~\ref{Fig2}--\ref{Fig4}), then we obtain the following two positive solutions: 
\begin{align}
r_{M}(a) &= \frac{K(a)+ \sqrt{K(a)^2-4AL}}{2A},  \label{EqB.8} \\
r_{m}(a) &= \frac{K(a)- \sqrt{K(a)^2-4AL}}{2A},  \label{EqB.9}
\end{align}
which are the values of $r$ at the intersections of the curve $y^*(r)$ and the dotted line $y^*=k$ in Figs.~\ref{Fig3} and \ref{Fig4}.
For the parameter sets in Figs.~\ref{Fig2} and \ref{Fig3}, $r_m(a)=0.91$, which is fully consistent with the first inverse bifurcation point $\tilde{r}_1 \simeq 0.91$. However, $r_M(a)=1.99$, which is significantly different from the second bifurcation point $r_2\simeq 1.65$, as seen in Fig.~\ref{Fig3}.

Since $r_M(a)>r_2$ and $\tilde{r}_1\geq r_m(a)$, at least for the parameter sets in Figs.~\ref{Fig3}--\ref{Fig5}, the width of the valley of the population $W$ is given by
\begin{align}
W = R - V_{\rm R} - V_{\rm L}, 
\end{align}
where
\begin{align}
R &\equiv r_M(a)-r_m(a) = \frac{\sqrt{K(a)^2-4AL}}{A}, \\
V_{\rm R} &\equiv r_M(a)-r_2,\\
V_{\rm L} &\equiv \tilde{r}_1-r_m(a).
\end{align}
The definition of $V_{\rm L}$ is based on that $\tilde{r}_1$ and $r_m(a)$, in general, have different values to each other.
Since the derivative of $R$ by $a$ is negative:
\begin{align}
\frac{\partial R}{\partial a} &=\frac{-K(a)}{\sqrt{K(a)^2-4AL}}<0,
\end{align}
then as $a$ increases, $R$ decreases. That is, $W$ is smaller if $V_{\rm L}$ and $V_{\rm R}$ are both non-decreasing function of $a$, which was confirmed numerically as illustrated in Fig.~\ref{Fig4}.

\section{Fixed-point analysis for the model without facultative predation}

The model without facultative predation ($H(r)=0$) is described by
\begin{align}
\frac{dx}{dt}&=-(a+d\cdot x)x+(c-r)\left.\frac{xy}{b+y}\right.,\label{EqB1}\\
\frac{dy}{dt}&=ry\left(1-\frac{y}{k}\right)+(c-r)\left.\frac{xy}{h+x}\right.,\label{EqB2}
\end{align}
and the Jacobian $A'$ is
\begin{align}
A' = \begin{bmatrix}- a - 2 d \cdot x + \dfrac{y \left(c - r\right)}{b + y} & - \dfrac{x y \left(c - r\right)}{\left(b + y\right)^{2}} + \dfrac{x \left(c - r\right)}{b + y}\\ - \dfrac{x y \left(c - r\right)}{\left(h + x\right)^{2}} + \dfrac{y \left(c - r\right)}{h + x} & r \left(1 - \frac{y}{k}\right) + \dfrac{x \left(c - r\right)}{h + x} - \dfrac{r y}{k}\end{bmatrix}.
\label{B2}
\end{align}

\subsection{Local stability of the trivial fixed points}\label{AppendixC1}
The Jacobian $A'_{(0,0)}$ of the first trivial fixed point $\vec{P}_0=(x,y)=(0,0)$ is 
\begin{align}
A'_{(0,0)}= \begin{bmatrix}- a & 0\\0 & r \end{bmatrix},\label{EqAB8}
\end{align}
and the Jacobian $A_{(0,k)}$ of another trivial fixed point $\vec{P}_a=(x,y)=(0,k)$ is 
\begin{align}
A'_{(0,k)} = \begin{bmatrix} - a + \dfrac{k \left(c - r\right)}{b + k} & 0 \\ \dfrac{k \left(c - r\right)}{h} & - r \end{bmatrix}.\label{EqAB9}
\end{align}
Using (\ref{EqAB8}) and (\ref{EqAB9}), we can obtain the local stability conditions for $\vec{P}_0$ and $\vec{P}_a$, which are same as in \ref{AppendixA1}.

\subsection{Local stability of the internal fixed point}\label{AppendixC2}

The equation for the $x$-nullcline is 
\begin{align}
x=\frac{1}{d}\left \{(c-r)\frac{y}{b+y}-a\right \}.
\end{align}
Assuming again that $b$ is sufficiently smaller than $y$, the $x$-coordinate of the fixed point is 
\begin{align}
x^* =  \frac{c-r-a}{d}.\label{B4}
\end{align}
Substituting this into the equation for the $y$-nullcline, the $y$-coordinate of the fixed point
is 
\begin{align}
y^*=\left.\frac{k\left \{rh d+c(c-r-a) \right \}}{r(h d+c-r-a)}\right..
\label{EqB8}
\end{align}
When $c-r>a$, $x^*$ and $y^*$ are both positive, and therefore,
\begin{align}
\vec{P}^{(n)}_I=(x^*,y^*)=\left(\frac{c-r-a}{d},\frac{k\left\{rh d+c(c-r-a)\right\}}{r\left(h d+c-r-a\right)}\right)\label{EqB9}
\end{align}
is the internal fixed point.

By substituting $\vec{P}^{(n)}_I$ into the Jacobian $A'$ in Eq.~(\ref{B2}) and assuming again that $b$ is sufficiently smaller than $y$, we have 
\begin{align}
A'_{(x^*,y^*)} \simeq \begin{bmatrix} -a  - 2d \cdot x^*+(c - r) & 0\\
-  \dfrac{x^* y^* \left(c - r\right)}{\left(h + x^*\right)^{2}} + \dfrac{y^* \left(c - r\right)}{h + x^*} & r  + \dfrac{x^* \left(c - r\right)}{h + x^*} - 2 \dfrac{r y^*}{k} \end{bmatrix}.\label{B17}
\end{align}
By substituting $x^*$ and $y^*$ in Eq.~(\ref{EqB9}) into Eq.~(\ref{B17}), the determinant and the trace can be derived as
\begin{align}
\operatorname{det}A'_{(x^*,y^*)}&=\frac{\left(c-r-a\right)\left \{rh d+c\left(c-r-a\right)\right \}}{h d+c-r-a},\\
\operatorname{Tr}A'_{(x^*,y^*)}&=a-c+\left.\frac{\left(c-r-a\right)\left(r-c\right)}{h d+c-r-a}\right..
\end{align}
Since $c-r>a$, then $\operatorname{det}A'_{(x^*,y^*)}>0$ and $\operatorname{Tr}A'_{(x^*,y^*)}<0$. Thus,  $\vec{P}^{(n)}_I$ is always locally stable.

\subsection{Proof that there are no closed orbits}\label{AppendixC3}

Let $g=1/xy$. Then, we can derive
\begin{equation}
\begin{aligned}
\frac{\partial }{\partial x}\left(g \dot{x}\right)+\frac{\partial }{\partial y}\left(g \dot{y}\right)
&=\frac{\partial }{\partial x}\left [\frac{1}{xy}\left \{ -\left(a+d\cdot x\right)x+\left(c-r\right)\frac{xy}{b+y}\right \}\right ] \\
&\qquad +\frac{\partial }{\partial y}\left[\frac{1}{xy} \left \{ ry\left(1-\frac{y}{k}\right)+\left(c-r\right)\frac{xy}{h+x}\right \}\right] \\
&=-\frac{d}{y}-\frac{r}{kx} \\
&< 0.
\end{aligned}
\end{equation}
Since the region $x,y>0$ is simply connected, the function $g$ and the system (\ref{EqB1}) and (\ref{EqB2}) satisfy the required smoothness conditions. Therefore, the model without facultative predation, Eqs.~(\ref{EqB1}) and (\ref{EqB2}), has no closed orbits in the first quadrant because of Dulac's criterion.

\bibliographystyle{elsarticle-harv}
\bibliography{Nakai_EcolModel_v1.bib}
\end{document}